\documentclass[showpacs,aps,prd,reprint,superscriptaddress,nofootinbib,longbibliography]{revtex4-2}

\usepackage[dvipdfmx]{graphicx}
\usepackage{amsmath,amssymb,bm,color,longtable,mathrsfs,amsfonts,slashed,ulem}
\usepackage[T1]{fontenc}
\usepackage[utf8]{inputenc}

\usepackage[colorlinks=true, pdfstartview=FitV, linkcolor=magenta,citecolor=blue, urlcolor=magenta,
bookmarks=true, bookmarksnumbered=true, breaklinks]{hyperref}

\allowdisplaybreaks

\newcommand \beq{\begin{eqnarray}}
\newcommand \eeq{\end{eqnarray}}

\newcommand{\tr}{{\rm tr}}
\newcommand{\Tr}{{\rm Tr}}

\newcommand{\Nc}{N_{\rm c}}
\newcommand{\Nf}{N_{\rm f}}

\newcommand{\vp}{ {\bm p}}

\newcommand{\la}{\langle}
\newcommand{\ra}{\rangle}

\newcommand{\calL}{\mathcal{L}}

\newcommand{\calS}{\mathcal{S}}

\newcommand{\calD}{\mathcal{D}}

\newcommand{\calM}{\mathcal{M}}

\newcommand{\calK}{\mathcal{K}}

\newcommand{\rmd}{\mathrm{d}}
\newcommand{\rmi}{\mathrm{i}}
\newcommand{\rme}{\mathrm{e}}

\newcommand{\bE}{ \bar{E}  }

\begin{document}
\begin{flushright}
\end{flushright}

\title{Statistical repulsion on hyperons in two-color dense QCD
}

\author{Masato Nagatsuka}
\email{ masato.nagatsuka.r4@dc.tohoku.ac.jp }
\affiliation{ Department of Physics, Tohoku University, Sendai 980-8578, Japan}

\author{Toru Kojo}
\email{ torukojo@post.kek.jp }
\affiliation{ Theory Center, IPNS, High Energy Accelerator Research Organization (KEK), 1-1 Oho, Tsukuba, Ibaraki, 305-0801, Japan }
\affiliation{ Graduate Institute for Advanced Studies, SOKENDAI, 1-1 Oho, Tsukuba, Ibaraki, 305-0801, Japan }

\date{\today}

\begin{abstract}
We investigate the onset of hyperons in baryonic (diquark) matter in two-color QCD (QC$_2$D) by introducing heavy quark doublets that emulate strange quarks. An even number of flavors is required to avoid the sign problem in lattice Monte Carlo simulations. 
To explore QC$_2$D matter containing both light and heavy quarks, 
we construct a model in which quarks interact with light-light, light-heavy (hyperonic), and heavy-heavy diquarks via Yukawa couplings.
As the quark chemical potential increases, the light diquarks condense first and form baryonic matter, and this onset density can be understood in hadronic terms. In contrast, the onset density of hyperons is substantially higher than that estimated from the hadronic sector of the model. This shift reflects an effective repulsion among baryons induced by the pre-occupied light quarks. The Pauli blocking of light quarks suppresses the attractive diquark correlations responsible, in vacuum, for making hyperons lighter than the sum of the constituent light and heavy quark masses. Implications for three-color QCD are also briefly discussed.
\end{abstract}


\maketitle

\section{Introduction}
\label{sec:Introduction}

Baryonic matter in quantum chromodynamics (QCD) consists of composite particles and, as the baryon density ($n_B$) increases, 
is expected to transform into quark matter. In the dilute regime, hadrons serve as natural effective degrees of freedom, allowing for reliable computations of the properties of dense matter \cite{Akmal:1998cf,Togashi:2017mjp,Drischler:2019xuo,Tews:2018kmu,Drischler:2021bup}.
Ab-initio nuclear calculations and statistical analyses of neutron star observations indicate that purely nucleonic computations remain valid only up to baryon densities of roughly $1.5$-$2 n_0$ ($n_0 \simeq 0.16, {\rm fm}^{-3}$, the nuclear saturation density) \cite{Tews:2018kmu,Drischler:2021bup,Brandes:2023bob}. 
Beyond this regime, various medium effects---such as the enhancement of many-body forces, the appearance of hadrons beyond nucleons, 
and structural modifications of hadrons---are expected to play an important role \cite{Masuda:2012kf,Masuda:2012ed,Kojo:2014rca,Baym:2017whm,Ma:2019ery,Marczenko:2022jhl}.

One of the important problems in neutron star physics is the so-called {\it hyperon softening problem}, or {\it hyperon puzzle} \cite{Lonardoni:2014bwa,Ye:2024meg,LiAng:2024xrl,Tolos:2020aln,Tolos:2016hhl,Chatterjee:2015pua,Togashi:2016fky,Miyatsu:2015kwa,Sun:2022yor,Sun:2018tmw}.
Typical hadronic calculations predict that hyperons appear at baryon densities around $n_B = 2$-$3 n_0$, but their emergence softens the equation of state (EOS) too much, rendering it incompatible with the well-established two-solar-mass constraints on neutron stars \cite{Demorest:2010bx,Fonseca:2021wxt,Antoniadis:2013pzd,TheLIGOScientific:2017qsa}.
One possible resolution is to introduce short-range repulsive interactions among hyperons and nucleons, such as $YN$ and $YNN$ forces, which delay the appearance of hyperons. Two-body $YN$ repulsion alone is typically insufficient, necessitating the inclusion of three-body $YNN$ repulsion \cite{Lonardoni:2014bwa}. Many-body calculations incorporating these interactions, with strengths constrained by $YN$ scattering data \cite{Li:2016paq} and hypernuclear spectroscopy \cite{Jinno:2025vgm,Le:2024rkd,Gal:2016boi,Friedman:2022bpw,Hiyama:2025yty,Miwa:2025adw,Hiyama:2019kpw} (for a recent review, see Ref.~\cite{Haidenbauer:2025zrr}), suggest that the resulting EOS can be compatible with the two-solar-mass constraints \cite{Tong:2025fzv,Tong:2025sui,Muto:2024upf,Gerstung:2020ktv}.

However, scenarios based on many-body repulsion raise questions regarding the convergence of the many-body expansion. When two- and three-body repulsive forces are comparable, it is natural to expect that four-, five-, and higher-body forces may also become important. Moreover, within the hadronic picture, it is not guaranteed that many-body effects always manifest as repulsion.
We also note that extrapolating $N$-body ($N>2$) repulsive contributions to the energy density, $\varepsilon_{\rm N-body} \sim n_B^N$, to high densities eventually leads to a squared sound speed $c_s^2 \sim N-1$, which violates the causality constraint $c_s^2 \le 1$ \cite{Kojo:2025vcq}. To study the global behavior of many-body forces at high density, analyses based on more fundamental degrees of freedom are essential.

Recently, there has been growing interest in quark-level descriptions of baryons in dense matter \cite{McLerran:2018hbz,Jeong:2019lhv,Duarte:2020xsp,Duarte:2020kvi,Zhao:2020dvu,Fukushima:2020cmk,Kojo:2021ugu,Kojo:2021hqh,Fujimoto:2023mzy,Tajima:2024qzj,Yamamoto:2024mta,Saito:2025yld,Gao:2024jlp,Gao:2025okn,Bentz:2025rla,Gartlein:2025zhd,Ivanytskyi:2025cnn}.
Since baryons are composed of quarks, the quark states as fermions must gradually become saturated as the baryon density increases. This phenomenon, referred to as {\it quark saturation} \cite{Kojo:2021hqh}, is expected to occur first at low momenta. As the density rises, the saturated region extends to higher momenta, eventually forming a quark Fermi sea.
Recent studies suggest that quark saturation can occur at $n_B \sim 1$-$3n_0$ \cite{Koch:2024qnz,McLerran:2024rvk,Nikolakopoulos:2025fgw}, even before baryons spatially overlap, 
thereby imposing nontrivial constraints on baryon dynamics. In the context of the hyperon puzzle, 
quark saturation induces statistical repulsion among nucleons and hyperons, 
since the pre-occupied light quarks disfavor the formation of hyperons containing up and down quarks \cite{Fujimoto:2024doc}.

It is important to test the idea of statistical repulsion induced by quark degrees of freedom, as it provides a more fundamental mechanism than descriptions based on many-body forces. Direct tests from first principles are unfortunately not feasible, mainly because lattice QCD simulations of dense systems suffer from the notorious sign problem.
However, simulations in two-color QCD (QC$_2$D) \cite{Takahashi:2009ef,Iida:2020emi,Itou:2025vcy,Iida:2022hyy,Murakami:2022lmq,Iida:2019rah,Iida:2020emi,Astrakhantsev:2020tdl,Astrakhantsev:2018uzd,Bornyakov:2020kyz,Boz:2018crd,Bornyakov:2017txe,Braguta:2016cpw,Braguta:2015zta,Braguta:2014gea,Buividovich:2020gnl} and in isospin QCD (QCD$_I$) \cite{Abbott:2023coj,Abbott:2024vhj,Brandt:2022hwy,Brandt:2017oyy} at finite chemical potentials are feasible. Indeed, by comparing schematic model calculations with lattice results, both macroscopic and microscopic insights have been obtained.
In this paper, we focus on dense QC$_2$D and discuss how this framework can be exploited to explore the concept of statistical repulsion.

Introducing strange quarks into QC$_2$D is formally straightforward; 
however, adding only a single flavor reintroduces the sign problem in lattice simulations, 
since the resulting Dirac determinant is no longer positive definite. 
To avoid this issue, we consider a heavy quark doublet, $Q_u$ and $Q_d$, with equal masses $M_Q$ \cite{Sakai:2025hrj}.
A hyperon analogous to $\Lambda (uds)$, $\Sigma_0 (uds)$, and similar states can then be constructed as a bound state of a light-heavy diquark. 
For simplicity, we focus on the isosinglet diquarks $u$–$Q_d$ and $d$–$Q_u$, with masses denoted by $m_Y$. 
In QC$_2$D, light diquarks are degenerate with the pion due to the Pauli-G\"{u}rsey symmetry. 
For hyperons, we expect $m_Y$ to be of the order of the kaon mass or roughly the sum of the constituent light and heavy quark masses.

In this model setup, the physics at low quark chemical potential $\mu$ is identical to that of two-flavor QC$_2$D. 
At $\mu = m_\pi/2$, the light diquarks condense, initiating the formation of baryonic matter. 
Near this threshold, the system can be well described as a Bose-Einstein condensate (BEC) of light diquarks. 
As $\mu$ increases, the BEC regime gradually evolves into the Bardeen-Cooper-Schrieffer (BCS) regime, where diquark pairs coexist on top of the quark Fermi sea. 
(For reviews on the BEC-BCS crossover, see, e.g., Refs.~\cite{schrieffer1999theory,Leggett_book,BCS-BEC_Parish}.)

Our central question concerns the behavior of hyperons when the baryon chemical potential, $\mu_B = 2\mu$, exceeds the hyperon mass $m_Y$. Neglecting interactions or constraints from the quark substructure, hyperons would appear at $\mu_B = m_Y$. In three-color QCD, such an onset softens the EOS, giving rise to the hyperon problem.
We argue that including light-quark loops in the effective potential generates an effective repulsion for hyperons, thereby substantially increasing the critical chemical potential and mitigating the hyperon-induced softening. This finding is consistent with results from the IdylliQ model \cite{Fujimoto:2023mzy} for quarkyonic matter \cite{McLerran:2007qj}.
In this paper, we compute the critical chemical potential for hyperon emergence by examining the sign of the quadratic terms in the hyperon fields. The EOS after hyperon onset is computed elsewhere.

This paper is structured as follows.
In Sec.~\ref{sec:model} we define our effective model for QC$_2$D with light and heavy flavors.
In Sec.~\ref{sec:eff_potential} we construct the renormalized effective potential.
In Sec.~\ref{sec:onset} we delineate the onset of hyperons.
Sec.~\ref{sec:summary} is devoted to summary.

\section{Model}
\label{sec:model}

\subsection{Lagrangian}
\label{sec:lagrangian}

The Lagrangian for the light quark sector is
\cite{Strodthoff:2013cua,Strodthoff:2011tz,Adhikari:2018cea,Adhikari:2016eef,Chiba:2023ftg,Ayala:2023cnt}
\begin{align}
\mathcal{L}_q 
&= \bar{q} \big[ \rmi \slashed{\partial} + \mu \gamma^0 \big] q 
	+ \frac{1}{\, 2 \,} (\partial_\mu \sigma)^2 + \frac{1}{2} (\partial_\mu \pi^a)^2  \notag  \\
&\quad + \frac{1}{\, 2 \,} (\partial_\mu - 2 \rmi \mu \delta_{\mu 0}) D \cdot (\partial^\mu + 2 \rmi \mu \delta^{\mu 0}) D^* \notag  \\
&\quad - g \bar{q} \big[ \sigma + \rmi \gamma_5 \pi^a \tau^a \big] q  \notag  \\
&\quad - \frac{\, g \,}{\, 2 \,} \left[ D^* (q^T C \rmi \gamma_5 \lambda_2 \tau_2 q) + D (\bar{q} C \rmi \gamma_5 \lambda_2 \tau_2 \bar{q}^T) \right] \notag \\
&\quad - \frac{\, m_\phi^2 \,}{2} \big( \phi_a^2 + |D|^2 \big) 
	- \frac{\, \lambda \,}{\, 24 \,}\big( \phi_a^2 + |D|^2 \big)^2
	\notag \\
&\quad + h_\sigma \sigma 
\,,
\end{align}
where $q=(u,d)^T$ is the iso-doublet, and bosonic fields $\phi_a = (\sigma, \vec{\pi})$.
Here $D$ and $D^*$ are an iso-singlet diquark and antidiquark.
The mass ($m_\phi^2 < 0$) and Yukawa coupling ($g$) for mesons and diquarks are related by the Pauli-G\"{u}rsey symmetry.
The term $h_\sigma \sigma$ explicitly breaks the chiral symmetry.

To mimic the strangeness degrees of freedom, 
we introduce heavy quarks with the constituent masses of $\sim$ 500 MeV as counterparts of strange quarks.
But introducing only a single flavor causes the sign problem in lattice simulations.
To employ the lattice as a simulator free from the sign problem, 
we introduce these heavy quarks as (iso)doublets so that the fermion determinant becomes positive definite.
The Lagrangian is\footnote{A four-flavor linear sigma model 
with the $SU(8)$ Pauli-G\"{u}rsey symmetry
has been studied recently in Ref.~\cite{Sakai:2025hrj}.
In this work we do not use such symmetry to constrain the model parameters.
}
\begin{align}
& \mathcal{L}_Q
= \bar{Q} \big( \rmi \slashed{\partial} - M_Q + \mu \gamma^0 \big) Q \notag \\
&\quad + \frac{1}{\, 2 \,} (\partial_\mu - 2 \rmi \mu \delta_{\mu 0} ) D_Q \cdot (\partial^\mu + 2 \rmi \mu \delta^{\mu 0}) D_Q^* \notag \\
&\quad - \frac{\, g_Q \,}{2} \left[ D_Q^* (Q^T C \rmi \gamma_5 \lambda_2 \tau_2 Q) + D_Q (\bar{Q} C \rmi \gamma_5 \lambda_2 \tau_2 \bar{Q}^T) \right] \notag \\
&\quad - \frac{\, m_{D_Q}^2 \,}{2} D_Q D_Q^* - \frac{\,  \lambda_Q \,}{24} |D_Q D_Q^*|^2
\,,
\end{align}
where $Q=(Q_u,Q_d)^T$ is an iso-doublet, 
and $D_Q$ is a color- and iso-singlet $Q_u Q_d$-diquark.

Finally, we consider couplings between the light and heavy quark sectors.
A diquark made of a light and a heavy quark is regarded as a hyperon.
Writing iso-singlet light-heavy diquark as $Y_u \sim u Q_d$ and $Y_d \sim d Q_u$,
the Lagrangian is $\calL_Y = \sum_{i=u,d} \calL_Y^i$, where
\begin{align}
&\mathcal{L}_Y^i 
= \frac{1}{\, 2 \,} (\partial_\mu - 2i\mu \delta_{\mu 0}) Y_i \cdot (\partial^\mu + 2 \rmi \mu \delta^{\mu 0}) Y_i^*  \notag \\
&\quad - \frac{\, g_Y \,}{2} \big[ Y^* (f^T C \rmi \gamma_5 \lambda_2 \tau_2 f ) + Y (\bar{f} C \rmi \gamma_5 \lambda_2 \tau_2 \bar{f}^T) \big]_i \notag \\
&\quad - \frac{\, m_Y^2 \,}{2} Y_i Y_i^* - \frac{\, \lambda_Y \,}{\, 24 \,} \big( Y_i Y_i^* \big)^2
\notag \\
&\quad 
- \kappa_{\phi Y} |\phi| Y_i Y_i^* 
- \lambda_{\phi Y} \big( \phi_a^2 + D^2 \big)  Y_i Y_i^*
\,,
\end{align}
where $f_u = (u, Q_d)^T$ and  $f_d = (d, Q_u)^T$, and $|\phi | = \sqrt{\phi_a^2}$ is the chiral scalar field.

In practice, it is useful to redefine fermion fields as
$\psi = q$, $\psi_Q=Q$, and $\psi_f = f$ with the relations
\beq
\big( \psi^C, \psi_Q^C, \psi_f^C \big)
=  \lambda_2 \tau_2 C 
\big( \bar{q}^T, \bar{Q}^T, \bar{f}^T \big)
\,.
\eeq
For example,
$\psi_{uR}^C = C \bar{d}_G^T $
and $ u_{R}^T C \lambda_2 \tau_2 = \bar{\psi}_{dG}^C $.
In particular,
\begin{align}
 (q^T C \rmi \gamma_5 \lambda_2 \tau_2 q) 
 &= \bar{\psi}^C \rmi \gamma_5 \psi \,,
 \notag \\
 (\bar{q} C \rmi \gamma_5 \lambda_2 \tau_2 \bar{q}^T)
 &= \bar{\psi} \rmi \gamma_5 \psi^C \,,
\end{align}
where the color and flavor are diagonal for $\psi$-bilinear as in usual mesonic operators.
The same applies for the other doublets.
Using these expressions we do not have to deal with color- and flavor-matrices
and the relationship with isospin QCD can be made explicit.

In the Namub-Gor'kov bases, the Lagrangian takes the form\footnote{
To describe diquark condensate with zero momentum, 
$\Delta \sim q(-p) q(p) \sim \bar{\psi}_C (-p) \psi (p)$, 
it is convenient to take the Fourier transform of $\Psi(x)$ to be
$\Psi(p) = ( \psi(p), \psi_C (-p) )$.
Then, the Fourier transform of the $\Psi$ bilinear becomes
\beq
\bar{\Psi} (p) \left[
\begin{matrix}
 \slashed{p} - \Sigma + \mu \gamma^0  
& \Delta \rmi \gamma_5
\\
\Delta^* \rmi \gamma_5
& 
 \slashed{p} - \Sigma - \mu \gamma^0 
\end{matrix}
\right] \Psi (p) \,.
\eeq
This form is identical to the quark bilinear in QCD$_I$.
Hence the analytic results of QC$_2$D can be transferred to QCD$_I$ by
simply modifying the factor $\Nc$ and $\Nf$.
}
\begin{align}
\calL_q
&= \bar{\Psi} \left[
\begin{matrix}
\rmi \slashed{\partial} - \Sigma + \mu \gamma^0  
& \Delta \rmi \gamma_5
\\
\Delta^* \rmi \gamma_5
& 
- \rmi \slashed{\partial} - \Sigma - \mu \gamma^0 
\end{matrix}
\right] \Psi
\notag \\
&\quad + \frac{1}{\, 2 \,} (\partial_\mu \sigma)^2 + \frac{1}{2} (\partial_\mu \pi^a)^2  \notag  \\
&\quad + \frac{1}{\, 2 \,} (\partial_\mu - 2 \rmi \mu \delta_{\mu 0}) D \cdot (\partial^\mu + 2 \rmi \mu \delta^{\mu 0}) D^* \notag  \\
&\quad - \frac{\, m_\phi^2 \,}{2}\big( \phi_a^2 + |D|^2 \big)  - \frac{\, \lambda \,}{\, 24 \,} \big( \phi_a^2 + |D|^2 \big)^2 \notag \\
&\quad + h_\sigma \sigma + h_D ( D + D^*) \,,
\end{align}
with $\Psi = ( \psi, \psi^C )^T /\sqrt{2} $, $\Sigma = g (\sigma + \rmi \gamma_5 \pi^a \tau^a )$, and $\Delta = g D$.
The heavy quark part is
\begin{align}
\calL_Q
&= \bar{\Psi}_Q \left[
\begin{matrix}
\rmi \slashed{\partial} - M_Q + \mu \gamma^0  
& \Delta_Q \rmi \gamma_5
\\
\Delta_Q^* \rmi \gamma_5
& 
- \rmi \slashed{\partial} - M_Q - \mu \gamma^0 
\end{matrix}
\right] \Psi_Q
\notag \\
& \quad + \frac{1}{\, 2 \,} (\partial_\mu - 2 \rmi \mu \delta_{\mu 0}) D_Q \cdot (\partial^\mu + 2 \rmi \mu \delta^{\mu 0}) D_Q^* \notag  \\
& \quad - \frac{\, m_{D_Q}^2 \,}{2} D_Q D_Q^*
- \frac{\,  \lambda_Q \,}{24} |D_Q D_Q^*|^2\,,
\end{align}
with $\Psi_Q = ( \psi_Q, \psi_Q^C )^T /\sqrt{2} $ and $\Delta_Q = g_Q D_Q$.
Finally, for later convenience we decompose the light-heavy sector as 
$\mathcal{L}_Y = \mathcal{L}_Y^{0} + \mathcal{L}_Y^{\rm mix} $,
where
\begin{align}
\hspace{-0.1cm}
\mathcal{L}_Y^0
& = \frac{1}{\, 2 \,}
(\partial_\mu - 2 \rmi \mu \delta_{\mu 0}) Y_i \cdot (\partial^\mu + 2 \rmi \mu \delta^{\mu 0}) Y_i^* 
 \notag \\
&\quad 
- \frac{\, m_Y^2 \,}{2} Y_i Y_i^*
- \frac{\, \lambda_Y \,}{\, 24 \,} \big( Y_i Y_i^* \big)^2
 \,,
\end{align}
for the kinetic terms and self-couplings among $Y_u$ and $Y_d$,
and, for the coupling between fermion pairs and bosonic fields,
\begin{align}
\mathcal{L}_Y^{\rm mix}
& = - g_Y 
\big[ Y_i^* \big( \bar{\psi}_{Q_i}^C \rmi \gamma_5 \psi_i \big)
	+ Y_i \big( \bar{\psi}_i \rmi \gamma_5 \psi_{Q_i}^C \big)  \big] 
 \notag \\
&\quad ~~~~
- \kappa_{\phi Y} |\phi| Y_i Y_i^* 
- \lambda_{\phi Y} \big( \phi_a^2 + D^2 \big)  Y_i Y_i^*
\,,
\end{align}
where we used
\beq
\hspace{-0.5cm}
\bar{\psi}_{Q_i}^C \rmi \gamma_5 \psi_i
	= \bar{\psi}_{i}^C \rmi \gamma_5 \psi_{Q_i} \,,
~~~
\bar{\psi}_{Q_i} \rmi \gamma_5 \psi_i^C 
	= \bar{\psi}_i \rmi \gamma_5 \psi_{Q_i}^C \,.
\eeq

\subsection{Counter terms}
\label{sec:counter_terms}

We compute the effective potential up to single fermion loop
for which the UV divergences must be handled.
For the single fermion loop we have only to renormalize the boson self-energies and vertices.
In practice it is convenient to use the counter term formalism.
First we attach indices $B$ to the fields, masses, and couplings,
e.g., $\psi \rightarrow \psi_B$, $m \rightarrow m_B$, $g \rightarrow g_B$, and so on,
and then decompose them into the renormalized and counter terms.
Schematically,
$\calL (\Phi_B, g_B) = \calL (\Phi, g) + \calL^{c.t.} (\Phi, \delta g, \delta Z_\Phi)$,
where ($ \Phi_B, g_B$) are general bare fields and couplings,
($\Phi, g$) are renormalized fields and couplings,
and ($\delta g, \delta Z_\Phi$) are counter terms for couplings and field normalization.
Explicitly, the counter term Lagrangian is \cite{Adhikari:2018cea,Adhikari:2016eef}
\begin{align}
\calL_q^{c.t.}
&= \frac{\, \delta Z_\phi \,}{\, 2 \,} (\partial_\mu \sigma)^2 + \frac{\, \delta Z_\phi \,}{\, 2 \,} (\partial_\mu \pi^a)^2  \notag  \\
&\quad + \frac{\, \delta Z_\phi \,}{\, 2 \,} (\partial_\mu - 2 \rmi \mu \delta_{\mu 0}) D \cdot (\partial^\mu + 2 \rmi \mu \delta^{\mu 0}) D^* \notag  \\
&\quad - \frac{\, \delta m_\phi^2 \,}{2} \big( \phi_a^2 + |D|^2 \big)  - \frac{\, \delta \lambda \,}{\, 24 \,} \big( \phi_a^2 + |D|^2 \big)^2 \,,
\end{align}
for the light quark sector,
and
\begin{align}
\calL_Q^{c.t.}
&= \frac{\, \delta Z_{D_Q} \,}{\, 2 \,} (\partial_\mu - 2 \rmi \mu \delta_{\mu 0}) D_Q \cdot (\partial^\mu + 2 \rmi \mu \delta^{\mu 0}) D_Q^* \notag  \\
& \quad - \frac{\, \delta m_{D_Q}^2 \,}{2} D_Q D_Q^*
- \frac{\, \delta \lambda_Q \,}{24} |D_Q|^4
\,,
\end{align}
for the heavy quark sector,
and
\begin{align}
(\mathcal{L}_Y^{u+d} )^{c.t.} 
& = \frac{\, \delta Z_Y \,}{\, 2 \,} (\partial_\mu - 2i\mu \delta_{\mu 0}) Y_i \cdot (\partial^\mu + 2 \rmi \mu \delta^{\mu 0}) Y_i^*  \notag \\
& - \frac{\, \delta m_Y^2 \,}{2} Y_i Y_i^* 
- \frac{\, \delta \lambda_Y \,}{24} \big( Y_i Y_i^* \big)^2
\notag \\
&
- \delta \kappa_{\phi Y} |\phi| Y_i Y_i^* 
- \delta \lambda_{\phi Y} \big( \phi_a^2 + D^2 \big)  Y_i Y_i^*
\,.
\end{align}
These counter terms are to be determined in vacuum and define our model.

Computing fermion loops inevitably yields the quartic terms.
Using the counter terms,
It is possible to define our model so that the renormalized quartic couplings $\lambda_Q$ and $\lambda_{Y}$ vanish in vacuum
at a specifically chosen renormalization scale.
With such setup we simplify our notation and then the nonzero quartic couplings solely arise from medium effects.

\subsection{Mean-field propagators}
\label{sec:propagators}

In this work we consider the density interval
where $\sigma$, $D $, and $D^* $ condense
but the other bosonic fields, $\pi$, $D_Q$, $D_Q^*$, and $Y_{u,d}$ are vanishing.
With these ansatz we use our model to just examine the onset of ``hyperons'' $Y_{u,d}$.
Below we assume $\pi=0$ and use the effective mass and gap $M_q = g \la \sigma \ra $ and $\Delta = g \la D \ra$
to describe the mean-field propagators for light quarks.

The mean field propagators for light quarks in such bosonic condensates are given by
\beq
\calS_q (p)
= \left[ 
\begin{matrix}
S_{11} (p) & S_{12} (p) \\
S_{21} (p) & S_{22} (p) 
\end{matrix}
\right]
= \left[ 
\begin{matrix}
\la \psi \bar{\psi} \ra     & \la \psi \bar{\psi}^C \ra \\
\la \psi^C \bar{\psi} \ra & \la \psi^C \bar{\psi}^C \ra
\end{matrix}
\right]
\eeq
where the normal $(S_{11}, S_{22})$ and the anomalous $(S_{12}, S_{21})$ propagators are
\begin{eqnarray}
S_{11}(p) &=& \sum_{\xi={\rm p}, {\rm a}}S^\xi_{11}(p)\Lambda_\xi  \gamma^0\ , \nonumber\\
S_{12}(p) &=& \sum_{\xi={\rm p}, {\rm a}}S^\xi_{12}(p)\Lambda_\xi  \gamma_5\ , \nonumber\\
S_{21}(p) &=& \sum_{\xi={\rm p}, {\rm a}}S^\xi_{21}(p)\Lambda^C_\xi  \gamma_5\ , \nonumber\\
S_{22}(p) &=& \sum_{\xi={\rm p}, {\rm a}}S^\xi_{22}(p)\Lambda^C_\xi  \gamma^0\ ,  \label{SDec}
\end{eqnarray}
with
\begin{eqnarray}
S^{\xi}_{11} (p)
	&=& \rmi \left(\frac{|u_{\xi}({\bm p})|^2 \,}{\, p_0 - \eta_q^\xi \epsilon_{\xi}({\bm p}) \,} 
		+ \frac{\, |v_{\xi}({\bm p})|^2 \,}{\, p_0 + \eta_\xi \epsilon_q^{\xi}({\bm p}) \,} \right)\ ,\nonumber\\
S^{\xi}_{12}(p) 
	&=&- \rmi \left( \frac{\, u_{\xi}^*({\bm p}) v_{\xi}^*({\bm p}) \,}{\, p_0-\epsilon_q^{\xi}({\bm p}) \,}  
		-  \frac{\, u_{\xi}^*({\bm p}) v_{\xi}^*({\bm p}) \,}{\, p_0+ \epsilon_q^{\xi}({\bm p}) \,} \right) \ ,\nonumber\\
S^{ {\xi}}_{21}(p) 
	&=& \rmi \left(\frac{\, u_{\xi}({\bm p}) v_{\xi}({\bm p}) \,}{\, p_0-\epsilon_q^{\xi}({\bm p}) \,} 
		- \frac{\, u_{\xi}({\bm p}) v_{\xi}({\bm p}) \,}{\, p_0 + \epsilon_q^{\xi}({\bm p}) \,} \right)\ , \nonumber\\
S^{ \xi }_{22} (p) 
	&=& \rmi \left( \frac{\, |v_{\xi} ({\bm p})|^2 \,}{\, p_0- \eta_\xi \epsilon_q^{\xi} ({\bm p}) \,} 
		+ \frac{\, |u_{\xi} ({\bm p})|^2 \,}{\, p_0 + \eta_\xi \epsilon_q^{\xi} ({\bm p}) \,} \right)\,,
\label{SEach}
\end{eqnarray}
where we introduced $\eta_{\rm p} = +1$ and $\eta_{\rm a} = -1$.
In these expressions, we have defined the positive-energy and negative-energy projection operators $\Lambda_{\rm p}$ and $\Lambda_{\rm a}$ by
\begin{eqnarray}
\Lambda_{\xi} (\vp) &=&\gamma_0 \frac{\, E_{q} (\vp) \gamma_0+ \eta_\xi \big( M_q + {\bm \gamma} \cdot{\bm p}\big) \,}{2E_{q} (\vp) } \,.
\end{eqnarray}
with $E_{q} (\vp) = \sqrt{ {\bm p}^2+M_q^2}$, and $\Lambda_{{\rm p}({\rm a})}^C = \Lambda_{{\rm a}({\rm p})}$, and
\begin{eqnarray}
\epsilon_q^{\xi}({\bm p}) &=& \sqrt{(E_{q}- \eta_\xi \mu)^2+|\Delta|^2} \,, 
\end{eqnarray}
are the dispersion relations for quasiparticles. 
The factors $u_{\rm p}({\bm p}) $, $v_{\rm p}({\bm p}) $,  $u_{\rm a}({\bm p}) $, and $v_{\rm a}({\bm p})$ satisfy relations
\begin{eqnarray}
|u_{\xi}({\bm p}) |^2 &=& \frac{1}{\, 2 \,} \left(1+ \frac{\, E_{q} - \eta_\xi \mu \,}{\, \epsilon_q^{\xi}({\bm p}) \,} \right)\,,  \nonumber\\
|v_{\xi}({\bm p}) |^2 &=& \frac{1}{\, 2 \,} \left(1 - \frac{\, E_{q} - \eta_\xi \mu \,}{\, \epsilon_q^{\xi}({\bm p}) \,} \right)\,.
\end{eqnarray}
Finally, the heavy quark propagator is
\beq
\calS_Q (p)
= \left[ 
\begin{matrix}
S_Q^{11} (p) & 0 \\
0 & S_Q^{22} (p) 
\end{matrix}
\right] \,,
\eeq
where
\begin{align}
S_Q^{11} (p)
&= (S_Q^{11})^{\rm p} \Lambda_{\rm p}^Q \gamma^0
	+ (S_Q^{11})^{\rm a} \Lambda_{\rm a}^Q \gamma^0
	\,,
\notag \\
S_Q^{22} (p)
&= (S_Q^{22})^{\rm p} (\Lambda_{\rm p}^Q)^C \gamma^0
	+ (S_Q^{22})^{\rm a} (\Lambda_{\rm a}^Q)^C \gamma^0 
	\,,
\label{SEach_heavy}
\end{align}
with $( \Lambda_{\rm p/a}^Q )^C = \Lambda_{\rm a/p}^Q$
and the projectors are defined in the same way as those for light quarks.
The coefficients are
\begin{eqnarray}
(S_Q^{11})^{\xi} 
&=& \rmi \left( \frac{\, \Theta(E_Q - \eta_\xi  \mu) \,}{\, p_0 - \eta_Q^\xi \epsilon_{\xi}({\bm p}) \,} 
		+ \frac{\, \Theta( \eta_\xi \mu - E_Q ) \,}{\, p_0 + \eta_Q^\xi \epsilon_{\xi}({\bm p}) \,} \right)\ ,
\nonumber\\
(S_Q^{22})^{\xi} 
&=& \rmi \left( \frac{\, \Theta( \eta_\xi \mu - E_Q ) \,}{\, p_0- \eta_Q^\xi \epsilon_{\xi} ({\bm p}) \,} 
		+ \frac{\, \Theta(E_Q - \eta_\xi  \mu) \,}{\, p_0 + \eta_Q^\xi \epsilon_{\xi} ({\bm p}) \,} \right)\,,
\label{SEach}
\end{eqnarray}
where $\epsilon_Q^\xi = |E_Q - \eta_\xi \mu|$.
Especially, for $M_Q > \mu$, the expression is simplified as 
\begin{align}
 (S_Q^{11})^{\xi} (p) &= \frac{\, \rmi \,}{\, p_0 + \mu - \eta_\xi E_Q \,} 
 \,,
 \notag \\
 (S_Q^{22})^{\xi} (p) &= \frac{\, \rmi \,}{\, p_0 - \mu + \eta_\xi E_Q \,}  
 \,.
\end{align}
In the above expressions we omit the usual small imaginary part $\rmi \delta$
to save the space, but in actual computations we make replacement $p_0 \rightarrow p_0 + \rmi p_0 \delta$
to describe the correct boundary condition.

\section{Effective potential}
\label{sec:eff_potential}
 
We compute the effective potential for 
possible condensates $\Phi = (\sigma, \pi, D, D_Q)$ and $Y = ( Y_u, Y_d )$.
In the following we treat all fields as real, without loss of generality.
The effective action can be computed by shifting bosonic fields into the condensed and fluctuation parts,
$\Phi \rightarrow \Phi + \hat{\Phi}$, $Y \rightarrow Y + \hat{Y}$,
and by computing the 1PI graphs;
\beq
\rme^{\rmi \Gamma (\Phi, Y) }
= \int \calD F \calD \hat{\Phi} \calD \hat{Y} \, \rme^{\rmi \int_x \calL ( F; \Phi + \hat{\Phi}, Y + \hat{Y} ) } \,,
\eeq
where $F$ collectively denotes fermion fields $\psi$, $\psi_Q$, and their conjugates.

In the mean field effective action, the fluctuations $\hat{\Phi}$ and $\hat{Y}$ are neglected,
\beq
\rme^{\rmi \Gamma_{\rm MF} (\Phi, Y) }
= \int \calD F \, \rme^{\rmi  \int_x \calL ( F; \Phi , Y ) } \,.
\eeq
We treat the coupling between hyperon fields $Y$ and light-heavy quark fields as perturbation,
\begin{align}
\rme^{\rmi \Gamma_{\rm MF} (\Phi, Y) }
&\simeq
 \int \calD F \, \rme^{\rmi  \int_x \big[ \calL_q + \calL_Q + \calL_Y^{0} \big]  } 
\notag \\
& \hspace{1cm}
 \times \bigg( 1 + \rmi \int_x \calL_Y^{\rm mix} + \cdots 
\bigg)\,.
\end{align}
The first nonzero contributions arise from the second order perturbation.
This generates the self-energy terms for the $Y$ fields.

We note that the $\psi_Q$ fields do not yield the anomalous propagators, i.e.,
in the density interval of $D_Q=0$.
The self-energy term comes from the product of the normal propagators,
\beq
- \rmi \frac{\, \Pi_Y \,}{2} |Y_i|^2 
\equiv  (-\rmi g_Y)^2 (-1) \Tr\big[ S_q^{11} \rmi \gamma_5 S_Q^{22} \rmi \gamma_5 \big] |Y_i|^2 \,.
\eeq
where the trace runs over momenta, color- and Dirac-indices, but not over flavors.
We delineate this term in the next section.
 
The effective potential $V_{\rm MF} = - \Gamma_{\rm MF} $ is 
\begin{align}
V_{\rm MF}
= V_{\rm tree} + V_{ct} + V_{\rm loop} + V_{\rm pert}^Y \,.
\end{align}
Assuming $\pi=D_Q = 0$, writing $M_q = g\sigma$, $\Delta =g D$, and $\Delta_{Y_i} = g_Y Y_i$,
the tree level potential is
\begin{align}
V_{\rm tree}
&= \frac{\, m_\phi^2  \,}{2g^2} M_q^2 
+ \frac{\, m_\phi^2 - 4\mu^2 \,}{2g^2} \Delta^2 
+ \frac{\, \lambda \,}{\, 24g^4 \,} \big( M_q^2 + \Delta^2 \big)^2
\notag \\
& 
- \frac{\, h_\sigma \,}{g} M_q - \frac{\, 2 h_D \,}{g} \Delta
\notag \\
&
+ \frac{\, m_{Y}^2 - 4\mu^2 \,}{2g_Y^2} \Delta_{Y_i}^2
+ \frac{\, \lambda_Y \,}{\, 24 \,} \big( \Delta_{Y_i}^2 \big)^2
\notag \\
& 
+ \frac{\, \kappa_{\phi Y} \,}{\, g g_Y^2 \,} M_q \Delta_{Y_i}^2 
+ \frac{\, \lambda_{\phi Y} \,}{\, g^2 g_Y^2 \,} \big( M_q^2 + \Delta^2 \big) \Delta_{Y_i}^2
 \,. 
\end{align}
The counter terms are
\begin{align}
V_{ct}
& = \! \frac{\, \delta m_\phi^2 \,}{\, 2g^2 \,} M_q^2
\! + \! \frac{\, \delta m_\phi^2 - 4\mu^2 \delta Z_\phi \,}{\, 2g^2 \,} \Delta^2  
\! + \! \frac{\, \delta \lambda \,}{\, 24g^4 \,} \big( M_q^2 \!+\! \Delta^2 \big)^2
\notag  \\
&~~
+ \frac{\, \delta m_{Y}^2 - 4\mu^2 \delta Z_{Y} \,}{\, 2g_Y^2 \,}  \Delta_{Y_i}^2 
+ \frac{\, \delta \lambda_Y \,}{\, 24 g_Y^4 \,} \big( \Delta_{Y_i}^2 \big)^2
\notag \\
&~~
+ \frac{\, \delta \kappa_{\phi Y} \,}{\, g g_Y^2 \,} M_q \Delta_{Y_i}^2 
+ \frac{\, \delta \lambda_{\phi Y} \,}{\, g^2  g_Y^2 \,} \big( M_q^2 + \Delta^2 \big)  \Delta_{Y_i}^2
\,.
\end{align}
The fermion loop yields $(\Nc = \Nf =2)$,
\begin{align}
V_{\rm loop}
&= - \Nf \Nc \int_{\vp} \big[ \epsilon_q^{\rm p} (\vp) + \epsilon_q^{\rm a} (\vp) \big]
\notag \\
&\quad - \Nf \Nc  \int_{\vp} \big[\, | E_Q (\vp) - \mu | + | E_Q (\vp) + \mu| \, \big] \,,
\end{align}
with $E_Q (\vp) = \sqrt{ \vp^2 + M_Q^2 \,}$.
For $M_Q > \mu$, the $\mu$-dependence of the heavy quark contributions vanish,
$|E_Q-\mu| + |E_Q+\mu| = E_Q - \mu + E_Q + \mu = 2E_Q$.
We discuss the isolation of the UV divergence shortly.

Finally the perturbative contribution is
\beq
V_{\rm pert}^Y
= \frac{\, \Pi_Y \,}{\, 2g_Y^2 \,} \Delta_{Y_i}^2 \,.
\eeq
Below we first fix the counter terms and then move 
to computations of $\Pi_Y$.
Meanwhile the determination of the renormalized parameters
will be done in separate computations.

\subsection{Light quark sector}
\label{sec:light_quark}

We evaluate 
\begin{align}
V_{\rm loop}^q (\mu)
& = -\Nc \Nf \int_{\vp} \big[ \epsilon_q^{\rm p} (\vp) + \epsilon_q^{\rm a} (\vp) \big] 
\notag \\
&\hspace{-0.5cm}
 = V_q^R (\mu) + V_q^{(1)} (M_q,\Delta)  + \mu^2 V_q^{(2)} (M_q,\Delta)  \,,
\end{align}
where the last two terms are divergent,
\begin{align}
V_q^{(1)} 
& = - 2 \Nc \Nf \int_{\vp} \sqrt{ \vp^2 + M_q^2 + \Delta^2 }\,, 
\notag \\
V_q^{(2)} 
& = - \Nc \Nf \int_{\vp} \frac{\,  \Delta^2 \,}{\, ( \vp^2 + M_q^2 + \Delta^2 )^{3/2} \,}\,. 
\end{align}
Subtracting these terms from $V_q$, 
the twice subtracted potential $V_q^R$, defined by $V_q - V_q^{(1)} - V_q^{(2)}$, is UV finite.
The divergent $V_q^{(1)}$ and $V_q^{(2)}$ can be computed
analytically in the dimensional regularization \cite{Adhikari:2018cea,Adhikari:2016eef},
\begin{align}
V_q^{(0)}
&= \frac{\, \Nc \Nf \,}{\, (4\pi)^2 \,}
\bigg( \frac{1}{\, \epsilon \,} + \frac{\, 3 \,}{2} - \ln \frac{\, M_q^2 + \Delta^2 \,}{\Lambda^2} \bigg) \big( M_q^2 + \Delta^2 \big)^2 \,,
\notag \\
V_q^{(2)}
&= \frac{\, 2\Nc \Nf \,}{\, (4\pi)^2 \,}
\bigg( \frac{1}{\, \epsilon \,}  - \ln \frac{\, M_q^2 + \Delta^2 \,}{\Lambda^2} \bigg) \big( - 2\Delta^2 \big) \,,
\end{align}
where $\Lambda$ is a renormalization scale
at which the renormalized parameters in our model are defined.

We use the $\overline{ {\rm MS} }$ scheme to remove the UV divergences.
The counter terms are
\begin{align}
&\delta m_\phi^2 = 0 \,,~~~~
\delta Z_\phi = - \frac{\, 2\Nc \Nf g^2 \,}{\, (4\pi)^2 \,} \frac{1}{\, \epsilon \,} \,,
\notag \\
& \hspace{1cm}
 \delta \lambda 
= - \frac{\, 24\Nc \Nf g^4 \,}{\, (4\pi)^2 \,} \frac{1}{\, \epsilon \,} \,.
\end{align}
Then, the effective potential for the light quark sector is
\begin{widetext}
\begin{align}
V_{\rm MF}^q 
& = ( V_{\rm tree} + V_{ct} + V_{\rm loop}  \big)^q
\notag \\
&= 
V_q^R (\mu)
- \frac{\, h_\sigma \,}{g} M_q 
- \frac{\, 2 h_D \,}{g} \Delta
+ \frac{\, m_\phi^2  \,}{2g^2} \big( M_q^2 + \Delta^2 \big)
- \frac{\, 2\mu^2 \Delta^2 \,}{ g^2 } 
\bigg[\, 1 
	- g^2 \frac{\, 2 \Nc \Nf \,}{\, (4\pi)^2 \,} 
	 \ln \frac{\, M_q^2 + \Delta^2 \,}{\Lambda^2} 
\bigg]
\notag \\
&\hspace{1.5cm}
 + \frac{\, 1 \,}{\, 24 g^4 \,} 
\bigg[\, \lambda
	+ g^4 \frac{\, 24 \Nc \Nf \,}{\, (4\pi)^2 \,} 
	 \bigg( \frac{3}{2} - \ln \frac{\, M_q^2 + \Delta^2 \,}{\Lambda^2} \bigg) 
 \bigg] (M_q^2 + \Delta^2 )^2
\,.
\end{align}
\end{widetext}
Later we express the physical pion mass $m_\pi$
in terms of $m_\phi$ and the other parameters
and show the onset of diquark condensed phase 
to be $\mu^{\rm onset} = m_\pi/2$.

\subsection{Heavy quark sector}
\label{sec:heavy_quark}

Similarly, the heavy quark loop for $M_Q > \mu$ is
\begin{align}
V_{Q}
= \frac{\, \Nc \Nf \,}{\, (4\pi)^2 \,}
\bigg( \frac{1}{\, \epsilon \,} + \frac{\, 3 \,}{2} - \ln \frac{\, M_Q^2 \,}{\Lambda^2} \bigg) M_Q^4 \,.
\end{align}
Since $M_Q$ is not dynamical, this constant can be simply removed by the vacuum subtraction,
or can be neglected.
Hence we may write
\beq
V_{\rm MF}^Q = 0 \,,
\eeq
at sufficiently low densities with
$M_Q > \mu$ and $\Delta_Q=0$.

\subsection{Light-heavy quark sector}
\label{sec:light-heavy_quark}

\subsubsection{The structure of the potential}
\label{sec:structure_potential}

The light-heavy potential is
\beq
V^Y_{\rm MF}
= 
\frac{\, C_2^Y \,}{2g_Y^2}  \Delta_{Y_i}^2
+ \frac{\, \lambda_Y + \delta \lambda_Y \,}{\, 24 g_Y^4 \,} \big( \Delta_{Y_i}^2 \big)^2
\eeq
where the coefficient of the quadratic $\Delta_{Y_i}$ term is
\begin{align}
C_2^Y
& = m_{Y}^2 - 4\mu^2 
+ \frac{\, 2 \kappa_{\phi Y} \,}{\, g  \,} M_q
+ \frac{\, 2 \lambda_{\phi Y} \,}{\, g^2  \,} \big( M_q^2 + \Delta^2 \big) 
\notag \\
& + \delta m_{Y}^2 - 4\mu^2 \delta Z_{Y} 
+ \frac{\, 2 \delta \kappa_{\phi Y} \,}{\, g  \,} M_q 
+ \frac{\, 2 \delta \lambda_{\phi Y} \,}{\, g^2 \,} \big( M_q^2 + \Delta^2 \big) 
\notag \\
& + \Pi_Y (\mu)  \,.
 \label{eq:V^Y_MF}
\end{align}
Below we evaluate the self-energy at zero momenum,
\begin{align}
&\Pi_Y (\mu) 
 = 2 \rmi g_Y^2 \Tr\big[ S_q^{11} (p) \rmi \gamma_5 S_Q^{22} (p) \rmi \gamma_5 \big] 
\notag \\
& = 2 \rmi g_Y^2 \Nc \sum_{\xi, \xi'= {\rm p, a} }
\int_p \tr\big[ \Lambda_\xi \gamma_0 \rmi \gamma_5 (\Lambda^Q_{\xi'})^C \gamma_0 \rmi \gamma_5 \big]
S_{11}^\xi (S_{22}^Q )^{\xi'} 
\,,
\end{align}
where $\Lambda_\xi = \Lambda_\xi (\vp) $, $\Lambda^Q_{\xi'} = \Lambda^Q_{\xi'} (\vp) $,
$S_{11}^{\xi} = S_{11}^{\xi} (p)$, and $(S_{22}^Q)^{\xi'} = (S_{22}^Q)^{\xi'} (p)$.
This function is UV divergent.
But its twice subtracted function $\Pi_Y^R$ can be made UV finite.
Then $\Pi_Y$ can be written as (the prime means $\mu^2$-derivative)
\beq
\hspace{-0.5cm}
\Pi_Y (\mu)
= \Pi_Y^R (\mu) + \Pi_{Y}^{(1)} (M_q,\Delta) + \mu^2 \Pi_{Y}^{(2)} (M_q,\Delta)  \,,
\label{eq:subtraction_functions}
\eeq
which defines $\Pi_Y^R$
with the divergent functions $\Pi_Y^{(1)}$ and  $ \Pi_Y^{(2)}$
to be determined shortly in Eq.~\eqref{eq:subtraction_constants}.
The divergences in $\Pi_Y^{(1)}$ and  $ \Pi_Y^{(2)}$ are cancelled by the counter terms.

First we carry out the $p_0$ integration,
\beq
\rmi \calM^{\xi \xi'} (\vp;\mu) = \int_{p_0} S_{11}^\xi (S_{22}^Q )^{\xi'} \,.
\eeq
For $M_Q > \mu$, it yields ($\epsilon_Q^\xi \equiv | E_Q - \eta_\xi \mu |$)
\beq
\calM^{\xi \xi}
= \frac{\, | u_{\xi} (\vp) |^2 \,}{\, \epsilon_q^{\xi} (\vp) + \epsilon_Q^{\xi} (\vp) \,} \,,~~~(\xi = \xi')
\eeq
for $\xi = \xi'$ corresponding to the particle-particle and antiparticle-antiparticle contributions,
and 
\beq
\calM^{\xi \xi' }
= \frac{\, | v_{\xi} (\vp) |^2 \,}{\, \epsilon_q^{\xi} (\vp) + \epsilon_Q^{\xi'} (\vp) \,} \,,~~~(\xi \neq \xi')
\eeq
for $\xi \neq \xi'$ representing the particle-antiparticle contributions.

Next we compute the trace of $\gamma$-matrices,
\begin{align}
\calK^{\xi \xi'}
&= \tr\big[ \Lambda_\xi \gamma_0 \rmi \gamma_5 (\Lambda^Q_{\xi'})^C \gamma_0 \rmi \gamma_5 \big]
\notag \\
&= 1 + \eta_\xi \eta_{\xi'} \frac{\, \vp^2 + M_q M_Q \,}{ E_q E_Q }\,,
\end{align}
which is $\mu$-independent. 

Now the self-energy is
\begin{align}
&\Pi_Y (\mu) 
 = - 2 g_Y^2 \Nc \sum_{\xi,\xi'} \int_{\vp} \calK^{\xi \xi'} (\vp) \calM^{\xi \xi'}  (\vp;\mu) 
  \,.
\end{align}
Before its renormalization, this (divergent) function is negative definite,
since both $\calK^{\xi \xi'}$ and $\calM^{\xi \xi'}$ are positive definite.
As we see later, however, the vacuum subtraction changes the sign.

\subsubsection{The sign of the self-energy}
\label{sec:sign_self}

Before proceeding to the renormalization of the self-energy for general $\Delta$,
it is useful to examine the sign of the self-energy holding $(M_q, \Delta) $ fixed to the vacuum values, $(M_0, 0)$.
Fixing the values of dynamical gaps makes the structure of the self-energy very transparent, as we see shortly.
The self-energy contains the integrands
\begin{align}
& \calK^{ {\rm pp} } \calM^{ {\rm pp} } + \calK^{ {\rm aa} } \calM^{ {\rm aa} } 
\notag \\
& = \calK^{ {\rm pp} }
\bigg[ \frac{\, \Theta(E_q - \mu) \,}{\, E_q + E_Q - 2 \mu \,} + \frac{\, 1 \,}{\, E_q + E_Q + 2 \mu \,}  \bigg]
\notag \\
& = \calK^{ {\rm pp} }
\bigg[ - \frac{\, \Theta(\mu - E_q) \,}{\, E_q + E_Q - 2 \mu \,}
\notag \\
& \hspace{1.5cm}
 + \frac{\, 2 \,}{\, E_q + E_Q \,} + \frac{\, 8\mu^2 \,}{\, ( E_q + E_Q )^3\,} + \cdots 
\bigg] \,,
\end{align}
and (reminder: $E_Q > \mu$)
\begin{align}
& \calK^{ {\rm pa} } \calM^{ {\rm pa} } + \calK^{ {\rm ap} } \calM^{ {\rm ap} } 
= \calK^{ {\rm pa} }
\frac{\, \Theta(\mu - E_q) \,}{\, E_Q - E_q \,} \,.
\end{align}
The terms without the step function depends on $\mu$ 
only through $M_q$ and $\Delta$;
these terms can be completely eliminated by the counter terms 
$ \delta m_{Y}^2$ and $ \delta Z_{Y}$, 
as we are holding $M_q$ and $\Delta$ fixed to the vacuum value. 
The resultant self-energy contributions are positive,
\begin{align}
& \Pi_Y (\mu) + \delta m_{Y}^2 - 4\mu^2 \delta Z_{Y} 
\notag \\
& = 2 g_Y^2 \Nc \int_{\vp} \Theta(\mu - E_q) 
\bigg[
\frac{\, \calK^{\rm pp} \,}{\, E_q + E_Q - 2 \mu \,}
- \frac{\, \calK^{\rm pa} \,}{\, E_Q - E_q \,}
\bigg] 
\notag \\
& > 2 g_Y^2 \Nc \int_{\vp} \Theta(\mu - E_q) 
\calK^{\rm pp} \bigg[
\frac{\, 1\,}{\, E_q + E_Q - 2 \mu \,}
- \frac{\,1 \,}{\, E_Q - E_q \,}
\bigg] 
\notag \\
& = 2 g_Y^2 \Nc \int_{\vp} \Theta(\mu - E_q) 
\calK^{\rm pp}
\frac{\, 2(\mu-E_q) \,}{\, (E_q + E_Q - 2\mu)(E_Q - E_q) \,}
\notag \\
& > 0 \,,
\end{align}
where $\calK^{\rm pp} > \calK^{\rm pa}$ is used.
Thus, the effective energy of the $Y$-diquarks increases.

This expression can be analytically estimated in 
the case with $M_Q \gg \mu \, (> M_q) $ as
\begin{align}
& \Pi_Y + \delta m_{Y}^2 - 4\mu^2 \delta Z_{Y} 
\notag \\
& \simeq \frac{\, 4 g_Y^2 \Nc \,}{\, M_Q \,} \int_{\vp} \Theta(\mu - E_q)  
= \frac{\, 4 g_Y^2 \Nc n_B \,}{\, M_Q \,} \,.
\end{align}
We note that $\calK^{\rm pp}-\calK^{\rm pa}\simeq2$ is justified when $|\vp|\ll M_q$.
The effective energy of $Y$ grows with $n_B$;
this represents an effective repulsion between $Y$ and light quarks.
If we retain the momentum dependence in the denominator,
the growth of the repulsion becomes milder,
with powers of $n_B$ less than the linear.

Below we generalize these considerations
by including the in-medium modifications of $M_q$ and $\Delta$.

\subsubsection{The structure of the UV divergence}
\label{sec:structure_divergence}

To isolate the UV divergence and apply the dimensional regularization,
we express the integral in (inverse) powers of $\bE \equiv \sqrt{ \vp^2 + M_q^2 + \Delta^2}$.
Expanding in this way, the effective potential can be expressed manifestly
in powers of $M_q^2 + \Delta^2$ which respects the Pauli-G\"{u}rsey symmetry in the original Lagrangian (in the chiral limit).
It is useful to note $E_q^2 = \bE^2 - \Delta^2$ and $E_Q^2 = \bE^2 - \Delta^2 + (M_Q^2 - M_q^2)$.

To specify the UV divergent terms,
we first examine the asymptotic behavior of $\calK^{\xi \xi'}$.
It is 
\beq
\!\!\!
\calK^{\xi \xi'}
= 1 + \eta_{\xi} \eta_{\xi'} 
- \frac{\, \eta_{\xi} \eta_{\xi'} (M_Q - M_q)^2 \,}{2 \bE^2 } + O(\bE^{-4})\,.
\eeq
For $\xi = \xi'$, this factor approaches constant at large energy and hence the divergence is severe;
we need to compute the other factor $\calM^{\xi \xi'}$ to $O(1/\bE^4)$.
For $\xi \neq \xi'$, the leading behavior is $\sim 1/E_q^2$ so that we have only to compute $\calM^{\xi \xi'}$ to $O(1/\bE^2)$.

Next, we examine the asymptotic behavior of $\calM^{\xi \xi'}$.
We first note $\calK^{\rm pp} = \calK^{\rm aa}$ and $\calK^{\rm pa} = \calK^{\rm ap}$
so that we encounter the combination of $\calM^{\rm pp+aa} = \calM^{\rm pp} + \calM^{\rm aa}$
and $\calM^{\rm pa+ap} =\calM^{\rm pa} + \calM^{\rm ap}$.
Then it is manifest that each sum is symmetric for $\mu \leftrightarrow -\mu$
so that $\Pi_Y$ is a function of $\mu^2$.
We expand the energies 
\begin{align}
\epsilon_q^\xi
&\simeq \bE - \eta_\xi \mu + \eta_\xi \frac{\,  \mu \Delta^2 \,}{\, 2 \bE^2 \,} 
+ O(\bE^{-3}) \,,
\notag \\
\epsilon_Q^\xi
&\simeq  \bE - \eta_\xi \mu + \frac{\, M_Q^2 - M_q^2 - \Delta^2 \,}{\, 2 \bE \,} 
+ O(\bE^{-3}) 
\,.
\end{align}
We also expand the spinors as
\begin{align}
| u_\xi |^2
&\simeq 
1 - \frac{\, \Delta^2 \,}{\, 4 \bE^2 \,} 
+ O(\bE^{-3}) 
\,,
\notag \\
| v_\xi |^2
&\simeq 
\frac{\, \Delta^2 \,}{\, 4 \bE^2 \,}
+ O(\bE^{-3}) 
\,.
\end{align}
Combining all these, we find 
\begin{align}
\calM^{\rm pp+aa}
& \simeq
\frac{1}{\, \bE \,} - \frac{\, M_Q^2 - M_q^2 \,}{\, 4 \bE^3 \,} + \frac{\, \mu^2 \,}{\, \bE^3 \,} 
+ O(\bE^{-5})
\,,
\notag \\
\calM^{\rm pa+ap}
& \simeq
\frac{\, \Delta^2 \,}{\, 4 \bE^3 \,} 
+ O(\bE^{-5})
\,.
\end{align}
Then, we can identify the UV divergences 
from the particle-particle and antiparticle-antiparticle contributions 
for the product $\calK \calM$ as
\begin{align}
&\calK^{\rm pp} \calM^{\rm pp+aa}
\notag \\
&\simeq
\frac{2}{\, \bE \,} - \frac{\, M_Q (M_Q - M_q) \,}{\, \bE^3 \,} + \frac{\, 2 \mu^2 \,}{\, \bE^3 \,} 
+ O(\bE^{-5})
\end{align}
while it turns out that the particle-antiparticle contributions do not yield the UV divergent term in $\Pi_Y$;
\beq
\calK^{\rm pa} \calM^{\rm pa+ap} \simeq O(\bE^{-5}) \,.
\eeq
The divergent functions $\Pi_Y^{(1)}$ and $\Pi_Y^{(2)}$, which are to be used in Eq.~\eqref{eq:subtraction_functions}, are
\begin{align}
& \Pi_Y^{(1)}
=
4 g_Y^2 \Nc \int_{\vp} 
\bigg[ - \frac{1}{\, \bE \,} + \frac{\, M_Q (M_Q-M_q) \,}{\, 2 \bE^3 \,} \bigg]
 \,,
\notag \\
& \Pi_Y^{(2)}
 = 4 g_Y^2 \Nc \int_{\vp} \frac{\, -1\,}{\, \bE^3 \,} \,.
 \label{eq:subtraction_constants}
\end{align}
Each term can be analytically evaluated by the dimensional regularization.
We find
\begin{align}
 \int_{\vp} \frac{1}{\, \bE \,}
 &= \frac{\, M_q^2 + \Delta^2 \,}{\, 8 \pi^2 \,}
 \bigg[ - \frac{1}{\, \epsilon \,} - 1 + \ln \frac{\, M_q^2 + \Delta^2 \,}{\, \Lambda^2 \,} \bigg]
 \,,
 \notag \\
  \int_{\vp} \frac{1}{\, \bE^3 \,}
 &= \frac{\, 1 \,}{\, 4 \pi^2 \,}
 \bigg[ \frac{1}{\, \epsilon \,} - \ln \frac{\, M_q^2 + \Delta^2 \,}{\, \Lambda^2 \,} \bigg]
 \,.
\end{align}
Hence the divergent functions are
\begin{align}
&\Pi_Y^{(1)}  
= \frac{1}{\, \epsilon \,} \frac{\, g_Y^2 \Nc \,}{\, 2\pi^2 \,} \big[ \Delta^2 + M_q^2 - M_Q M_q + M_Q^2 \big]
\notag \\
&~~~
+ \frac{\, g_Y^2 \Nc\,}{\, 2\pi^2 \,} \big( \Delta^2 + M_q^2 \big) 
\notag \\
&~~~
- \frac{\, g_Y^2 \Nc \,}{\, 2\pi^2 \,} \big( \Delta^2 + M_q^2 - M_Q M_q + M_Q^2 \big) \ln \frac{\, \Delta^2 + M_q^2 \,}{\, \Lambda^2 \,} 
\,,
\notag \\
&\Pi_Y^{(2)} 
= \frac{\, g_Y^2 \Nc \,}{\, \pi^2 \,} 
\bigg[ - \frac{1}{\, \epsilon \,} 
	+ \ln \frac{\, M_q^2 + \Delta^2 \,}{\, \Lambda^2 \,} \bigg] \,.
\end{align}
Now the light-heavy effective potential can be renormalized.
In the $\overline{\rm MS}$ scheme
we set the counter terms as
\begin{align}
&\delta Z_Y = - \frac{\, g_Y^2 \Nc \,}{\, 4 \pi^2 \,} \frac{1}{\, \epsilon \,} \,,~~~~
\delta m_Y^2 = - \frac{\, g_Y^2 \Nc M_Q^2 \,}{\, 2 \pi^2 \,} \frac{1}{\, \epsilon \,} \,,
\notag \\
&
\delta \kappa_{\phi Y} = \frac{\, g g_Y^2 \Nc M_Q \,}{\, 4 \pi^2 \,} \frac{1}{\, \epsilon \,} \,,~~~~
\delta \lambda_{\phi Y}
= - \frac{\, g^2 g_Y^2 \Nc \,}{\, 4 \pi^2 \,} \frac{1}{\, \epsilon \,} \,.
\end{align}
Cancelling the divergences by counter terms,
the renormalized light-heavy potential \eqref{eq:V^Y_MF} now reads
\beq
V^Y_{\rm MF} 
= \frac{\, \big( C_2^Y \big)_R \,}{2g_Y^2}  \Delta_{Y_i}^2
+ \frac{\, \lambda_Y + \delta \lambda_Y \,}{\, 24 g_Y^4 \,} \big( \Delta_{Y_i}^2 \big)^2
\eeq
with the renormalized quadratic coefficient
\begin{align}
\big( C_2^Y \big)_R 
& = 
m_{Y}^2 - 4\mu^2 
+ \frac{\, 2 \kappa_{\phi Y} \,}{\, g  \,} M_q 
+ \frac{\, 2\lambda_{\phi Y} \,}{\, g^2 \,} \big( M_q^2 + \Delta^2 \big) 
\notag \\
& + \frac{\, g_Y^2 \Nc \,}{\, 2 \pi^2 \,} \big( M_q^2 + \Delta^2 \big) 
\notag \\
&
- \frac{\, g_Y^2 \Nc \,}{\, 2\pi^2 \,}  \big( \Delta^2 + M_q^2 - M_Q M_q + M_Q^2 \big) \ln \frac{\, M_q^2 + \Delta^2 \,}{\, \Lambda^2 \,} 
\notag \\
& + \mu^2 \frac{\, g_Y^2 \Nc \,}{\, 2 \pi^2 \,}  \ln \frac{\, M_q^2 + \Delta^2 \,}{\, \Lambda^2 \,} 
\notag \\
& + \Pi_Y^R (\mu)
 \,.
 \label{eq:eff_pot_full}
\end{align}
To determine $\delta \lambda_Y$, we also need to compute a four point function of fields $Y$.
We skip such computations, 
since we focus only on the coefficients of of the $\Delta_{Y}^2$ terms 
which determines the onset of $Y_{u,d}$,
provided that the appearance of $Y_{u,d}$ does not accompany the first order phase transition
($\Delta_{Y}$ smoothly increases from zero).

\section{The onset of hyperons}
\label{sec:onset}

\begin{figure}[t]
\vspace{-.3cm}
\begin{center}
\includegraphics[width=8.8 cm]{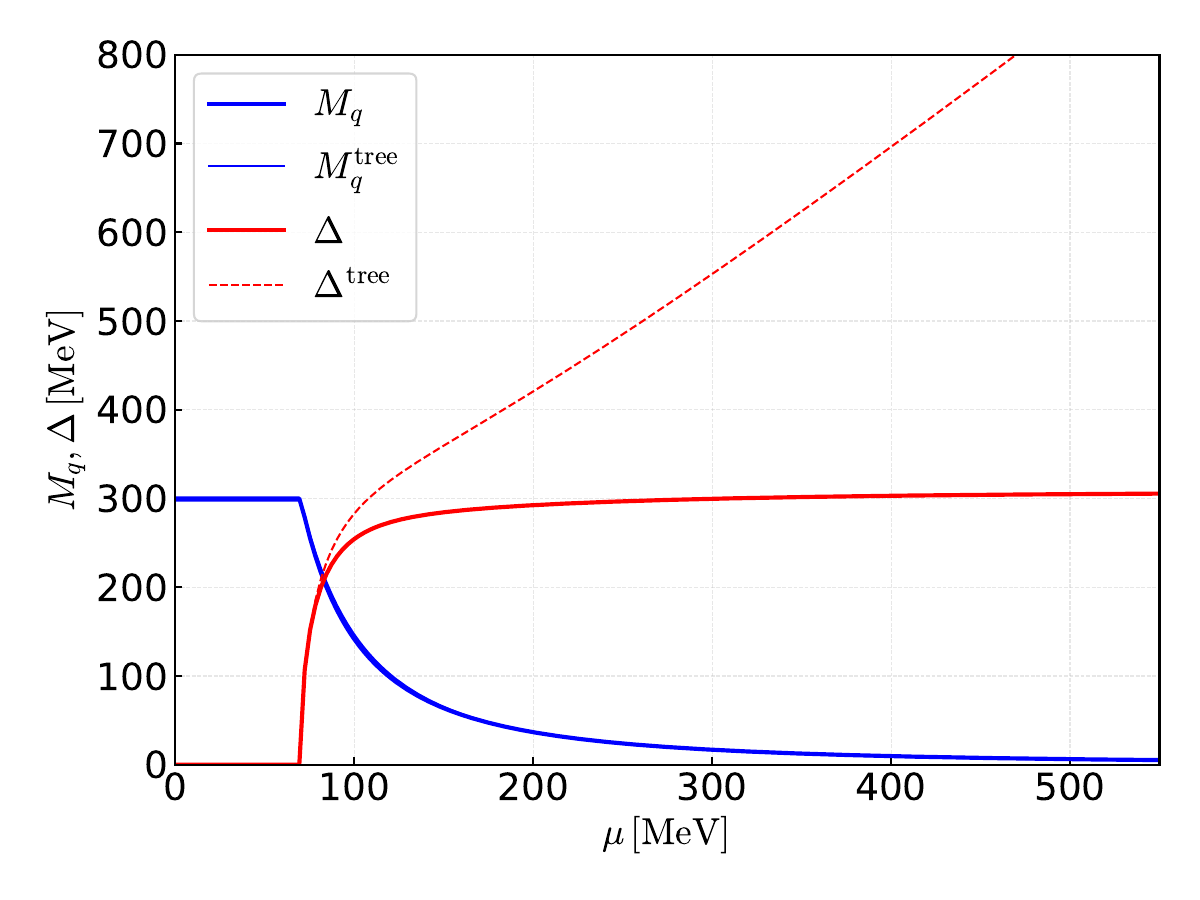}
\end{center}
\vspace{-.7cm}
\caption{The evolution of $(M_q, \Delta)$ as functions of $\mu$.
The onset of diquark condensate is $\mu = m_\pi/2$.
The $\Delta$ at tree level grows as $\Delta \sim \mu$,
while inclusion of the quark coupling tempers the growth, resulting $\Delta \sim M_0$.
$M_q$ and $M_q^{\rm tree}$ largely overlap and the difference is not visible.
}
\label{fig:gap_vs_mu}
\end{figure}   

\begin{figure}[t]
\vspace{-.3cm}
\begin{center}
\includegraphics[width=8.8 cm]{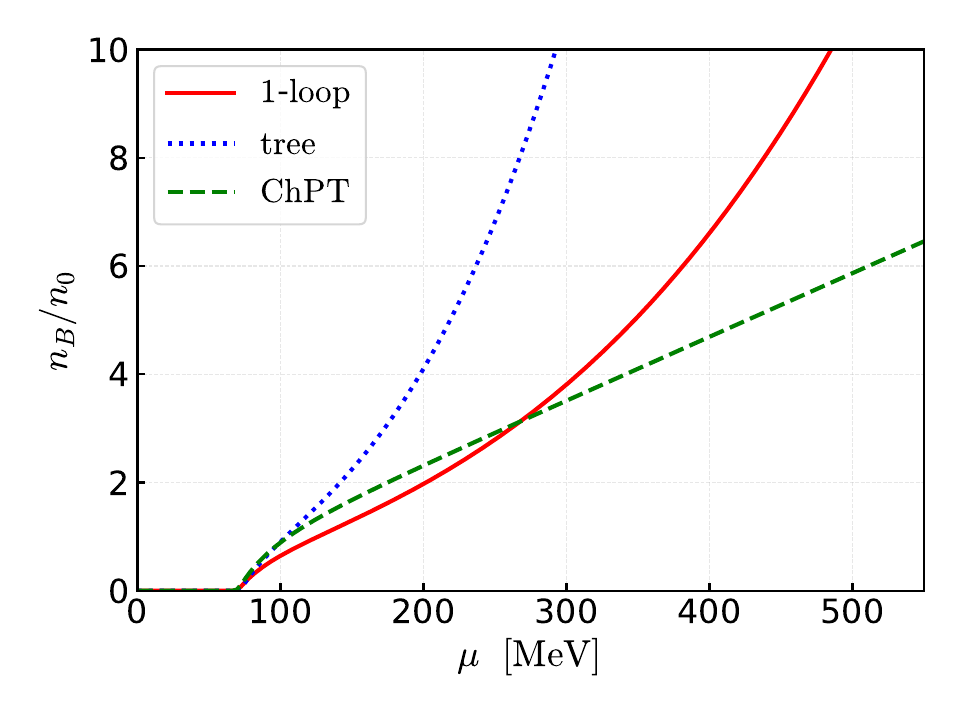}
\end{center}
\vspace{-.7cm}
\caption{The baryon density $n_B$ as functions of $\mu$
for the one-loop, tree, and ChPT cases.
We take $n_0 = 0.16\,{\rm fm}^{-3}$ as a unit.
}
\label{fig:n_vs_mu}
\end{figure}   


\begin{figure}[t]
\vspace{-.3cm}
\begin{center}
\includegraphics[width=8.8 cm]{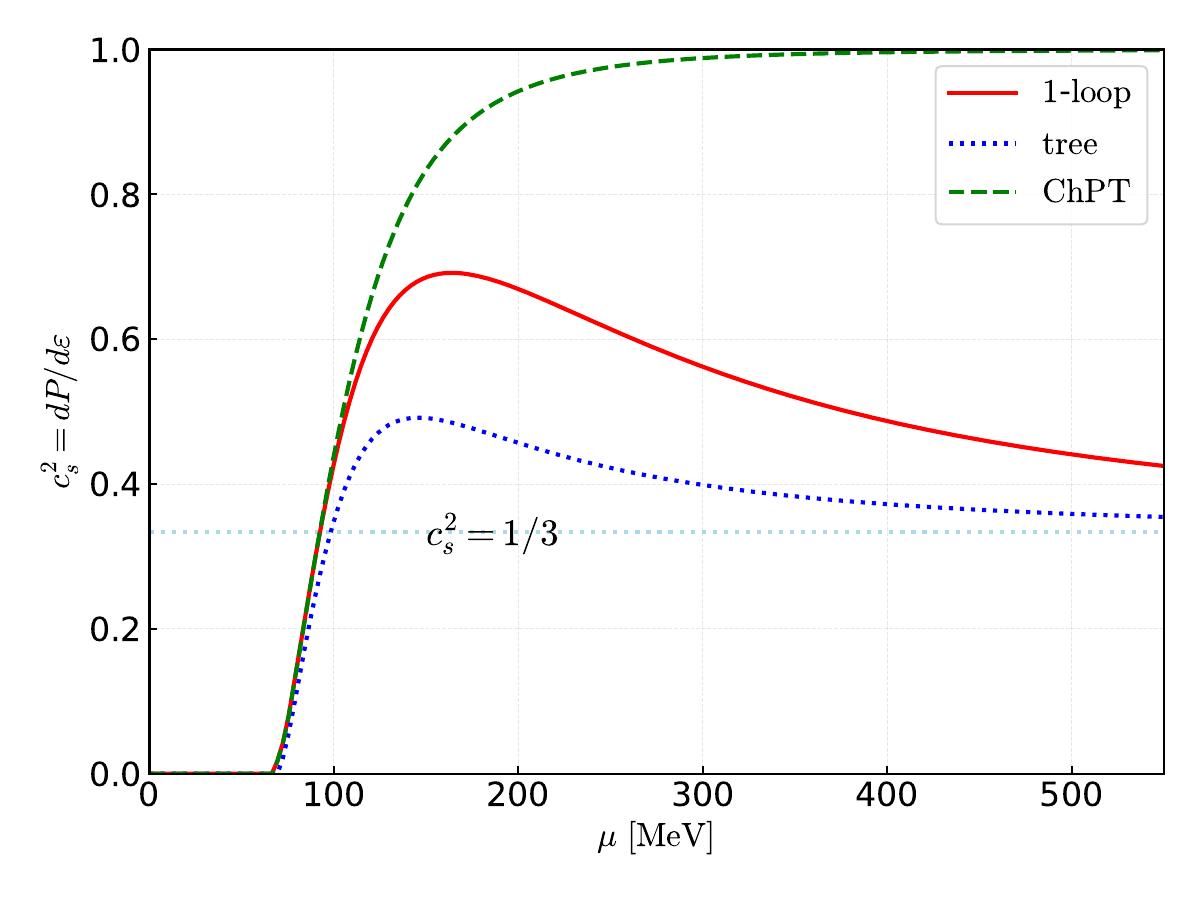}
\end{center}
\vspace{-.7cm}
\caption{The squared sound speed $c_s^2 = \rmd P/\rmd \varepsilon$ as functions of $\mu$
for the one-loop, tree, and ChPT cases.
The conformal limit $1/3$ is also plotted for eye-guide.
}
\label{fig:cs2_vs_mu}
\end{figure}   


From now on we read off the coefficients of $\Delta_{Y_i}^2$ terms in Eq.~\eqref{eq:eff_pot_full}.
We determine the critical chemical potential where the coefficients of  $\Delta_{Y_i}^2$ terms vanish.

Below we avoid details of hadronic interactions by setting $\kappa_{\phi Y}$ and $\lambda_{\phi Y}$ to zero.
We also take our renomalization scale $\Lambda$ to be $M_0$, the vacuum effective quark mass,
and all couplings should be regarded as those defined at $\Lambda =M_0$.
Then the renormalized vacuum mass of $Y$ is given by
\begin{align}
\big( m_Y^{\rm vac} \big)^2
 = m_Y^2 + \Pi_Y^R (\mu=0) + \frac{\, g_Y^2 \Nc M_0^2 \,}{\, 2\pi^2 \,} 
 \,.
\end{align}
Eliminating the tree level mass $m_Y$ in favor of the renormalized vacuum mass $m_Y^{\rm vac}$,
the in-medium quadratic coeffcient of $Y$ can be written as
\begin{align}
&\hspace{-0.1cm}
\big( C_2^Y \big)_R 
= \big( m_Y^{\rm vac} \big)^2 - 4 \mu^2  
+ \big[ \Pi_Y^R (\mu) - \Pi_Y^R (0)  \big]
\notag \\
&
 + \frac{\, g_Y^2 \Nc \,}{\, 2\pi^2 \,}  \big(  \Delta^2 + M_q^2- M_0^2 \big)  
\notag \\
&
- \frac{\, g_Y^2 \Nc \,}{\, 2\pi^2 \,}  \big( \Delta^2 + M_q^2 - M_Q M_q + M_Q^2 \big) \ln \frac{\, M_q^2 + \Delta^2 \,}{\, M_0^2 \,}
\notag \\
& + \mu^2 \frac{\, g_Y^2 \Nc \,}{\, 2 \pi^2 \,}  \ln \frac{\, M_q^2 + \Delta^2 \,}{\, M_0^2 \,} \,.
\end{align}
Assuming a transition to the phase of condensed $Y$ is the second order type,
the $Y$-diquarks begin to condense at the chemical potential where the coefficient of the quadratic term vanishes.

\subsubsection{Model parameters}
\label{sec:model_parameters}

In our analyses we use the following set of parameters:
\begin{align}
& g \simeq 4.3\,,~~~~ \sqrt{-m_\phi^2} \simeq 510\, {\rm MeV}\,, 
\notag \\
& \lambda \simeq 137\,,~~~~ h_\sigma^{1/3} \simeq 112\, {\rm MeV} \,,
\end{align}
with which we obtain
\begin{align}
& M_0 \simeq 300\, {\rm MeV}\,,~~~~m_\sigma \simeq 545\, {\rm MeV} \,,
\notag \\
& m_\pi \simeq 140\, {\rm MeV}\,,~~~~ f_\pi \simeq 85\, {\rm MeV}\,.
\end{align}
For parameters involving heavy quarks, we use
\beq
M_Q = 500\, {\rm MeV} \,,~~~~ \kappa_{\phi Y} = \lambda_{\phi Y}=0\,.
\eeq
We take $g_Y/g =1.0$ and $m_Y^{\rm vac} = 500$ MeV as our reference values,
but vary them to delineate the effects of the statistical repulsion.

\subsubsection{Baryonic matter of light quarks}
\label{sec:gaps}

First we describe the evolution of a baryonic matter made of light quarks.
Here we switch off the hyperon and heavy quark sectors.
The equation of state of matter is given by
\beq
P (\mu) = - V_{\rm MF}^q ( \mu; M_q^*, \Delta^*)
\eeq
where $M_q^*$ and $\Delta^*$ are the solutions of the gap equations.
The quark number ($n$) and baryon number ($n_B$) densities are given by $n = \Nc n_B = \partial P/\partial \mu $.
The energy density is $\varepsilon = \mu n - P$.
The general trends have been studied in the previous works
so here we review only a few important points, see Ref.~\cite{Chiba:2023ftg,Kojo:2024sca} for more details.

We compare the results of purely hadronic model, i.e., tree level and chiral perturbation theory (ChPT) results,
to the one-loop results with the quark-diquark coupling.
The low energy theorem leads to the universal behaviors in the low density limit;
the deviation of these model results reflect the sensitivity to the hadron-hadron interactions and
the importance of the quark substructure.

Shown in Fig.~\ref{fig:gap_vs_mu} are the evolution of gaps,  $(M_q, \Delta)$, as functions of $\mu$,
at the one-loop and tree levels.
The light diquarks begin to condense at $\mu = m_\pi/2$
and the chiral effective quark mass begins to decrease accordingly.
The $\Delta$ grows rapidly just beyond the threshold.
At higher density the behaviors are very different in the purely hadronic model and
model with the quark-diquark coupling.
At tree level,\footnote{At tree level we have to re-tune model parameters
as they do not contain the quark loop effects.
We use the tree level relations
\begin{align}
& m_{\phi, {\rm tree}}^2 = - \frac{\, m_\sigma^2 - 3 m_\pi^2 \,}{2} \,,~~~~
\lambda_{\rm tree}  = \frac{\, 2(m_\sigma^2 - m_\pi^2) \,}{  \la \sigma \ra_{\rm tree}^2} \,,
\notag \\
& \la \sigma \ra_{\rm tree} \simeq 0.81 f_\pi 
\,,
\end{align}
with which the tree and one-loop results coincide for $\mu \le m_\pi/2$.
}
$\Delta$ grows as $\Delta \sim \mu$,
since, within the hadronic part,
the quadratic term $\sim -\mu^2 \Delta^2$ must be balanced 
by the quartic (repulsive) term $\sim \lambda \Delta^4$.
In contrast, with the quark-diquark couplings,
the quark loop generates a $\mu^2$ term which can balance
with the $\sim -\mu^2 \Delta^2 \ln \Delta^2 $ term;
hence the strong $\mu^2$ dependence is factored out
and the resulting $\Delta $ can be insensitive to $\mu$.
The asymptotic behavior of the pressure is
$P\sim C_0 \mu^4 + C_1 \mu^2 \Delta^2 + \cdots$
where $C_0$ and $C_1$ depend on the hadronic parameters only weakly;
the details of hadronic parameters affect the value of $\Delta$
but not the coefficients.
The bulk part is fixed by the quark descriptions.
See Ref.~\cite{Chiba:2023ftg,Kojo:2024sca} for more details.

Next we examine the baryon density $n_B$ as a function of $\mu$ (Fig.~\ref{fig:n_vs_mu}).
We take $n_0 = 0.16\, {\rm fm}^{-3}$ for our unit
as this unit is often used for three-color QCD.
We expect that the size of a baryon for two-color and three-color QCD do not differ significantly,
$\sim 0.5$-0.8 fm, then the overlap of diquarks in QC$_2$D should occur around $\sim 5n_0$
which corresponds to $\mu \simeq 350$ MeV.
The results for different models coincide near the threshold $\mu=m_\pi/2$ but
they soon begin to deviate at $\mu \sim$ 90-100 MeV.

In Fig.~\ref{fig:cs2_vs_mu}, we plot the squared sound speed $c_s^2 = \rmd P/\rmd \varepsilon$.
All models show the similar growth of $c_s^2$ just above the threshold.
Beyond $\mu \simeq$ 90-100 MeV,
the ChPT approaches $c_s^2=1$ which comes from the behavior 
$P\sim f_\pi^2 \mu^2$, which should be regarded as an artifact at very large $\mu$.
The other two models yield the sound speed peaks around $\mu\sim$ 150-200 MeV or $n_B \sim$ 1.5-2$n_0$
but with the different magnitude,
and then both relax to the conformal value $c_s^2 =1/3$
after $\mu^4$-term dominate the pressure.

\subsubsection{Onset of hyperons}
\label{sec:onset2}

\begin{figure}[t]
\vspace{-.3cm}
\begin{center}
\includegraphics[width=8.8 cm]{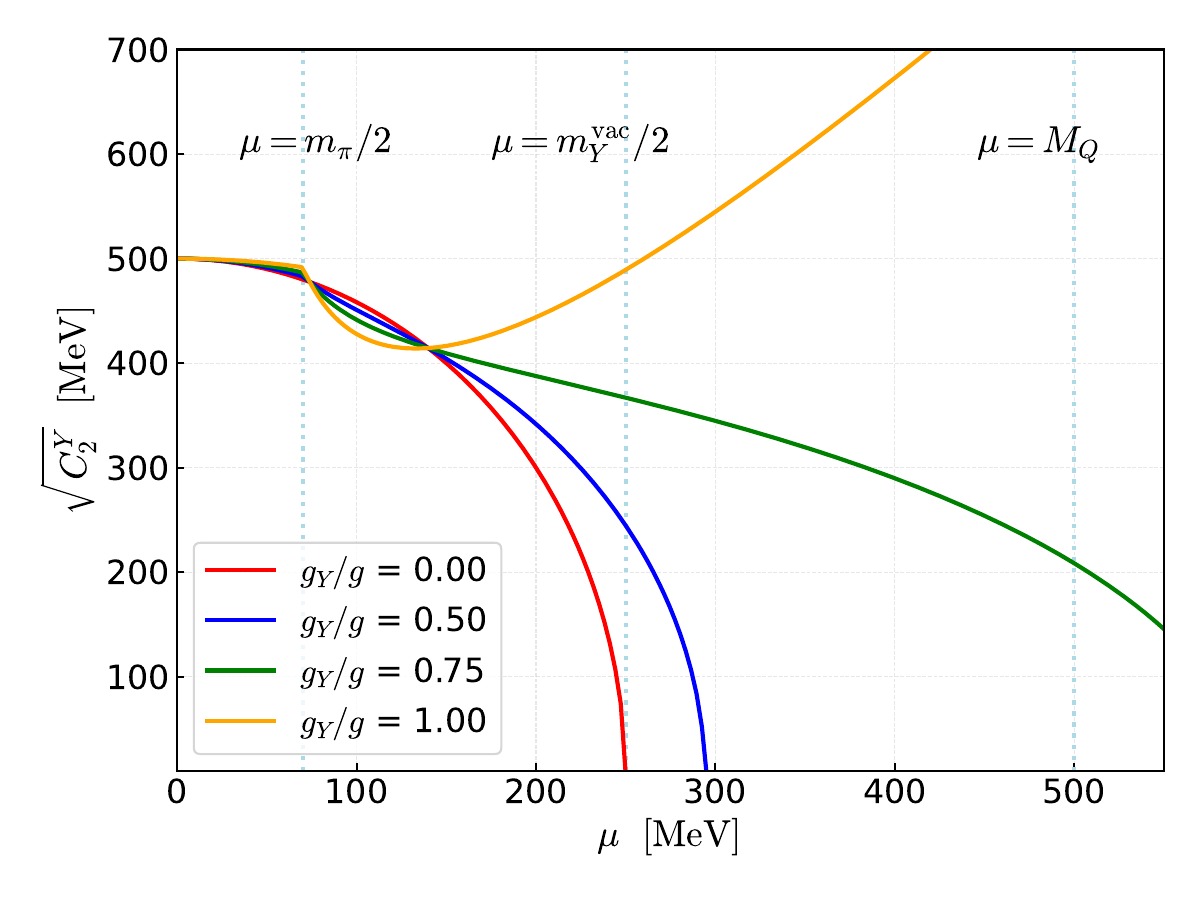}
\end{center}
\vspace{-.4cm}
\caption{ The density evolution of the coefficient $C_2^Y$ for the quadratic term of $Y_{u,d}$.
When $C_2^Y$ becomes negative, the hyperon fields condense.
At tree level the hyperon condensates emerge at $\mu = m_Y^{\rm vac} /2$ or $\mu_B = m_Y^{\rm vac} $.
After including the coupling $g_Y$ between quark and heavy-light diquark,
the onset chemical potential is shifted to a higher value for a greater $g_Y$.
}
\label{fig:c2Y}
\end{figure}   

\begin{figure}[th]
\vspace{-.3cm}
\begin{center}
\includegraphics[width=8.8 cm]{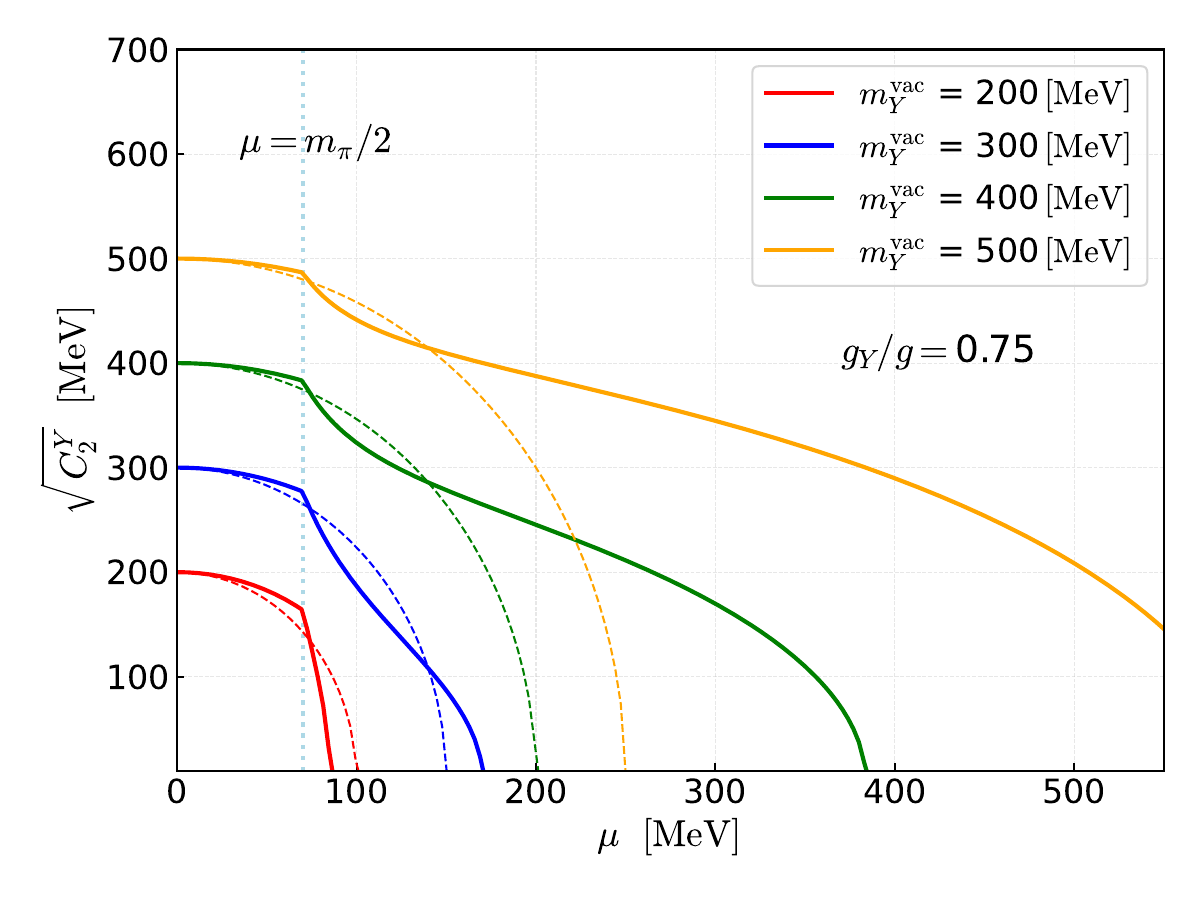}
\end{center}
\vspace{-.7cm}
\caption{ The density evolution of $C_2^Y$ for the vacuum hyperon mass, $m_Y^{\rm vac} = 200, 300, 400$, and 500 MeV,
with $g_Y/g=0.75$.
The dashed lines are the results for $g_Y=0$.
For a larger $m_Y^{\rm vac}$, more quark states are occupied so that
the statistical repulsion sets in before $Y$ condenses.
}
\label{fig:c2Y_mYvary_gY75}
\end{figure}   

\begin{figure}[th]
\vspace{-.3cm}
\begin{center}
\includegraphics[width=8.8 cm]{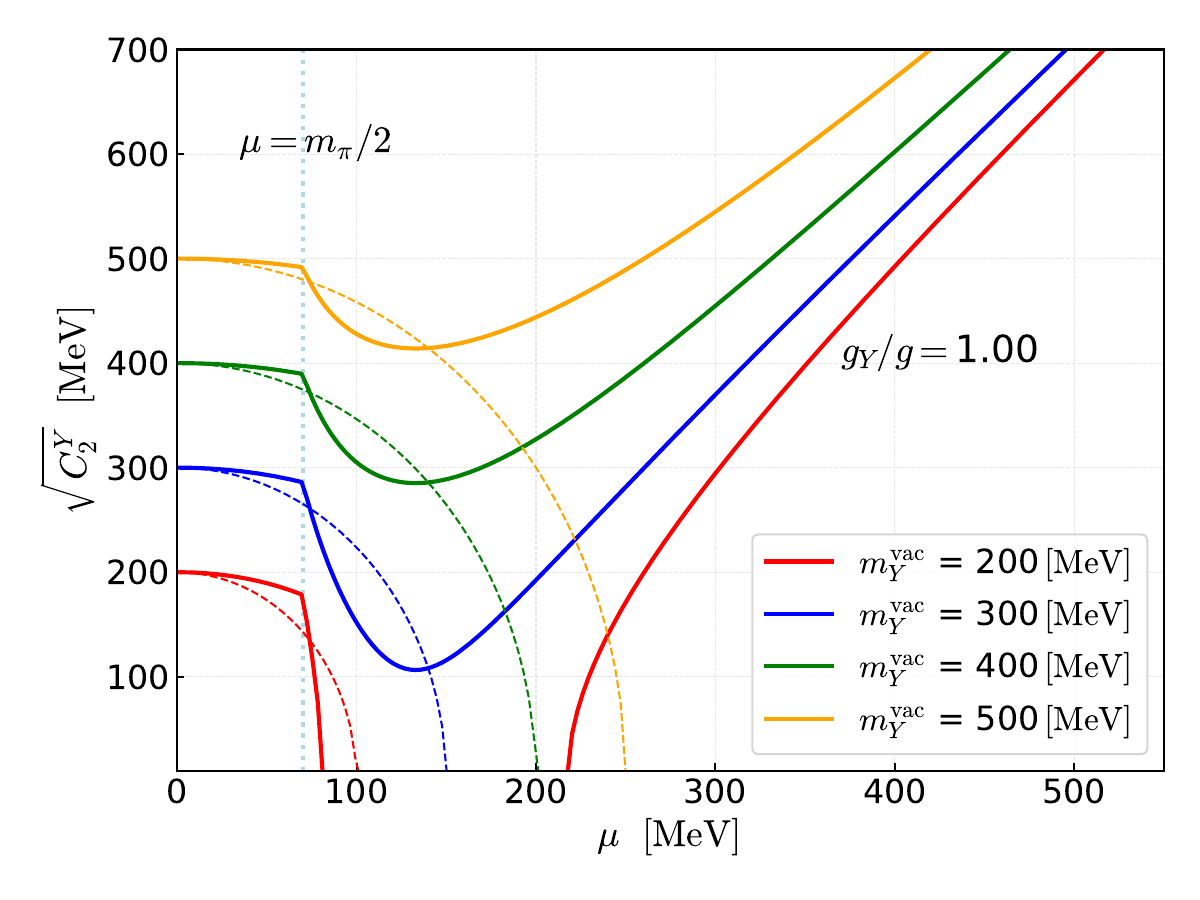}
\end{center}
\vspace{-.7cm}
\caption{ The same plot as Fig.~\ref{fig:c2Y_mYvary_gY100} except $g_Y/g =1.0$.
}
\label{fig:c2Y_mYvary_gY100}
\end{figure}   

Using the above mentioned density-dependent gaps,
we numerically evaluate the evolution of $C_2^Y$.

Shown in Fig.~\ref{fig:c2Y}
are $C_2^Y$ for various strength of the coupling $g_Y$ and $(m_Y^{\rm vac}, M_Q) = (500, 500)$ MeV.
At tree level including the hadronic part only,
$\sqrt{C_2^Y}$ drops from the vacuum value $m_Y^{\rm vac}$ to zero
at $\mu = m_Y^{\rm vac}/2$ or $\mu_B = m_Y^{\rm vac}$.
We recall that we are switching off the interactions among
light hadrons $(\sigma, \pi, D)$ and hyperons $Y$, i.e.,
setting $\kappa_{\phi Y} = \lambda_{\phi Y} = 0$,
so the onset is determined independently from the light hadron sector.

Including quark loops,
the hyperons $Y$ are influenced by a matter formed by light quarks.
The quark substructure effects in $Y$ become stronger for a larger $g_Y$;
explicitly, $g_Y$ characterizes the strength of the transition
\beq
Y ~\leftrightarrow~ qQ \,,
\eeq
so that with a greater $g_Y$ one has more chances to observe the composite structure
through the coupling of $qQ$ to the chemical potential.
The $g_Y=0$ corresponds to the tight-binding limit
and we only observe the elementary particle aspect of $Y$.
For more realistic considerations, 
it is natural to expect its strength to be comparable to the Yukawa coupling between a light quark and a light-light diquark.
In particular, if we assume $g_Y = g$, hyperons do not condense 
before the quark chemical potential reaches the heavy quark mass, $\mu=M_Q$.

Whether the statistical repulsion becomes important or not depends
on how early light quark states are occupied.
This can be seen in Figs.~\ref{fig:c2Y_mYvary_gY75} for $g_Y/g=0.75$ and \ref{fig:c2Y_mYvary_gY100} for $g_Y/g=1.0$,
where $m_Y^{\rm vary}$ is varied from 200 MeV to 500 MeV.
If $m_Y^{\rm vac}$ is very close to $m_\pi$,
the critical chemical potential $\mu = m_Y^{\rm vac}/2$ is reached 
before the light quarks substantially occupy the low momentum states.
In this case the shift of the critical chemical potential is small.

Some comments deserve for the case with $m_Y^{\rm vac} = 200$ MeV and $g_Y/g = 1.0$;
here $Y$ condenses at $\mu < m_Y^{\rm vac}/2$ but the condensates melt around $\mu \simeq 220$ MeV.
The condensation below the tree level estimate $\mu < \mu^{\rm tree} = m_Y^{\rm vac}/2$ is not regarded as generic; 
just above $\mu = m_\pi/2$, the details of the chiral restoration and diquark formation are as important as the statistical repulsion.
At higher $\mu$, the trend of $M_q$ and $\Delta$ becomes stable,
and the self-energy of $Y$ is mainly determined by the statistical repulsion.
In particular, for $g_Y/g=1.0$ and $m_Y^{\rm vac} = 200$ MeV, the repulsive effects which grow with density can destroy the condensates of $Y$.

\section{Summary}
\label{sec:summary}

We study the onset of ``hyperons'' in dense QC$_2$D by introducing heavy quarks that mimic strange quarks.
The introduction of the heavy doublet is motivated to avoid the sign problem 
in lattice Monte Carlo simulations.

QCD and QC$_2$D differ because baryons are fermions in the former but bosons in the latter.
Meanwhile, one can still study the statistical constraints due to quarks
in the same way for both cases.
In this respect, it should be useful to delineate the properties of matter with hyperons in QC$_2$D.

In dilute regime hyperons behave as if elementary particles in reaction to the increase of the chemical potential.
Once light diquarks condense, however, light quarks in hyperons are subject to the statistical constraint
and affect the hyperon self-energies.
If we drop off quark contributions by hand (tree level approximation), 
hyperons begin to condense at $\mu_B (= 2\mu) = m_Y^{\rm vac}$.
This trend changes by allowing hyperons to transform into quark intermediate states which are subject to the statistical constraints. 
A stronger coupling shifts the onset chemical potential for hyperons to a larger value.
For $g_Y = g$, we found that hyperons do not condense except $m_Y$ very close to $m_\pi$, 
and the strangeness enters the system not through composite hadrons, 
but through elementary strange quarks which appear for $\mu > M_Q$.

In this work we study hyperons up to the chemical potential just before the onset of $Y$ or heavy quarks $Q$, i.e., $\mu_B < m_Y^{\rm vac}$ or $\mu_B < 2M_Q$.
With this restriction, we could use light quark propagators in the background of $M_q$ and $\Delta$, 
while treated heavy quarks as impurities.
This allows us simple computations.
Meanwhile, it is also important to study how hyperons behave beyond their onset
and how the EOS softens.
We expect that the statistical repulsion caused by pre-occupied light quarks
not only delays the appearance of hyperons but also tempers the softening associated with their emergence.
The EOS after the emergence of hyperons can be computed
by using light and heavy propagators in the background of $M_q$, $\Delta$, and $Y$.
This will be discussed elsewhere.

\begin{acknowledgments}
T.K. thanks Profs. Wolfram Weise and Yuki Fujimoto for discussions about the neutron star constraints and nuclear theories.
This work was supported by 
JSPS KAKENHI Grant No. 23K03377 (TK);
JSPS Research Fellows No. JP24KJ0412 (MN),
and by the Graduate Program on Physics for the Universe (GPPU) at Tohoku university.
\end{acknowledgments}

\bibliography{ref}

@misc{Sakai:2025hrj,
    author = "Sakai, Manato and Suenaga, Daiki",
    title = "{Roles of $U(1)$ axial anomaly effects in cold and dense two-color QCD with $2+2$ flavors}",
    eprint = "2509.20468",
    archivePrefix = "arXiv",
    primaryClass = "hep-ph",
    month = "9",
    year = "2025"
}

@article{Kojo:2025vcq,
    author = "Kojo, Toru",
    title = "{Stiffening of matter in quark{\textendash}hadron continuity: A mini-review}",
    eprint = "2412.20442",
    archivePrefix = "arXiv",
    primaryClass = "nucl-th",
    doi = "10.1016/j.jspc.2025.100088",
    journal = "J. Subatomic Part. Cosmol.",
    volume = "4",
    pages = "100088",
    year = "2025"
}

@article{Brandes:2023bob,
    author = "Brandes, Len and Weise, Wolfram",
    title = "{Constraints on Phase Transitions in Neutron Star Matter}",
    eprint = "2312.11937",
    archivePrefix = "arXiv",
    primaryClass = "nucl-th",
    doi = "10.3390/sym16010111",
    journal = "Symmetry",
    volume = "16",
    number = "1",
    pages = "111",
    year = "2024"
}

@article{Akmal:1998cf,
    author = "Akmal, A. and Pandharipande, V. R. and Ravenhall, D. G.",
    title = "{The Equation of state of nucleon matter and neutron star structure}",
    eprint = "nucl-th/9804027",
    archivePrefix = "arXiv",
    doi = "10.1103/PhysRevC.58.1804",
    journal = "Phys. Rev. C",
    volume = "58",
    pages = "1804--1828",
    year = "1998"
}

@article{Tews:2018kmu,
    author = "Tews, Ingo and Carlson, Joseph and Gandolfi, Stefano and Reddy, Sanjay",
    title = "{Constraining the speed of sound inside neutron stars with chiral effective field theory interactions and observations}",
    eprint = "1801.01923",
    archivePrefix = "arXiv",
    primaryClass = "nucl-th",
    reportNumber = "INT-PUB-18-001, LA-UR-17-31455",
    doi = "10.3847/1538-4357/aac267",
    journal = "Astrophys. J.",
    volume = "860",
    number = "2",
    pages = "149",
    year = "2018"
}

@article{Drischler:2019xuo,
    author = "Drischler, Christian and Haxton, Wick and McElvain, Kenneth and Mereghetti, Emanuele and Nicholson, Amy and Vranas, Pavlos and Walker-Loud, Andr{\'e}",
    title = "{Towards grounding nuclear physics in QCD}",
    eprint = "1910.07961",
    archivePrefix = "arXiv",
    primaryClass = "nucl-th",
    reportNumber = "LLNL-JRNL-786701",
    doi = "10.1016/j.ppnp.2021.103888",
    journal = "Prog. Part. Nucl. Phys.",
    volume = "121",
    pages = "103888",
    year = "2021"
}

@misc{Haidenbauer:2025zrr,
    author = "Haidenbauer, Johann and Mei{\ss}ner, Ulf-G. and Nogga, Andreas",
    title = "{Ab initio description of hypernuclei}",
    eprint = "2508.05243",
    archivePrefix = "arXiv",
    primaryClass = "nucl-th",
    month = "8",
    year = "2025"
}

@article{Le:2024rkd,
    author = "Le, Hoai and Haidenbauer, Johann and Mei{\ss}ner, Ulf-G. and Nogga, Andreas",
    title = "{Light {\ensuremath{\Lambda}} Hypernuclei Studied with Chiral Hyperon-Nucleon and Hyperon-Nucleon-Nucleon Forces}",
    eprint = "2409.18577",
    archivePrefix = "arXiv",
    primaryClass = "nucl-th",
    doi = "10.1103/PhysRevLett.134.072502",
    journal = "Phys. Rev. Lett.",
    volume = "134",
    number = "7",
    pages = "072502",
    year = "2025"
}

@article{Gal:2016boi,
    author = "Gal, A. and Hungerford, E. V. and Millener, D. J.",
    title = "{Strangeness in nuclear physics}",
    eprint = "1605.00557",
    archivePrefix = "arXiv",
    primaryClass = "nucl-th",
    doi = "10.1103/RevModPhys.88.035004",
    journal = "Rev. Mod. Phys.",
    volume = "88",
    number = "3",
    pages = "035004",
    year = "2016"
}

@article{Friedman:2022bpw,
    author = "Friedman, E. and Gal, A.",
    title = "{Constraints from {\ensuremath{\Lambda}} hypernuclei on the {\ensuremath{\Lambda}}NN content of the {\ensuremath{\Lambda}}-nucleus potential}",
    eprint = "2204.02264",
    archivePrefix = "arXiv",
    primaryClass = "nucl-th",
    doi = "10.1016/j.physletb.2023.137669",
    journal = "Phys. Lett. B",
    volume = "837",
    pages = "137669",
    year = "2023"
}

@misc{Hiyama:2025yty,
    author = "Hiyama, E. and Doi, T.",
    title = "{Cluster phenomena using few-body and Lattice QCD theories}",
    eprint = "2511.14155",
    archivePrefix = "arXiv",
    primaryClass = "nucl-th",
    reportNumber = "RIKEN-iTHEMS-Report-25",
    month = "11",
    year = "2025"
}

@article{Miwa:2025adw,
    author = "Miwa, Koji and Nakazawa, Kazuma and Tamura, Hirokazu and Hiyama, Emiko and Takahashi, Toshiyuki",
    title = "{Nuclear systems with strangeness and baryon{\textendash}baryon interactions}",
    doi = "10.1140/epja/s10050-025-01571-z",
    journal = "Eur. Phys. J. A",
    volume = "61",
    number = "6",
    pages = "128",
    year = "2025"
}

@article{Hiyama:2019kpw,
    author = "Hiyama, E. and Sasaki, K. and Miyamoto, T. and Doi, T. and Hatsuda, T. and Yamamoto, Y. and Rijken, Th. A.",
    title = "{Possible lightest $\Xi$ Hypernucleus with Modern $\Xi N$ Interactions}",
    eprint = "1910.02864",
    archivePrefix = "arXiv",
    primaryClass = "nucl-th",
    reportNumber = "number:RIKEN-iTHEMS-Report-19,YITP-19-93",
    doi = "10.1103/PhysRevLett.124.092501",
    journal = "Phys. Rev. Lett.",
    volume = "124",
    number = "9",
    pages = "092501",
    year = "2020"
}

@misc{Jinno:2025vgm,
    author = "Jinno, Asanosuke and Haidenbauer, Johann and Mei{\ss}ner, Ulf-G.",
    title = "{Properties of hyperons in nuclear matter from chiral hyperon-nucleon interactions at next-to-next-to-leading order}",
    eprint = "2509.24459",
    archivePrefix = "arXiv",
    primaryClass = "nucl-th",
    month = "9",
    year = "2025"
}

@article{Li:2016paq,
    author = "Li, Kai-Wen and Ren, Xiu-Lei and Geng, Li-Sheng and Long, Bingwei",
    title = "{Strangeness $S=-1$ hyperon-nucleon scattering in covariant chiral effective field theory}",
    eprint = "1603.07802",
    archivePrefix = "arXiv",
    primaryClass = "hep-ph",
    doi = "10.1103/PhysRevD.94.014029",
    journal = "Phys. Rev. D",
    volume = "94",
    number = "1",
    pages = "014029",
    year = "2016"
}

@misc{Tong:2025fzv,
    author = "Tong, Hui and Elhatisari, Serdar and Mei{\ss}ner, Ulf-G. and Ren, Zhengxue",
    title = "{Multi-strangeness matter from ab initio calculations}",
    eprint = "2509.26148",
    archivePrefix = "arXiv",
    primaryClass = "nucl-th",
    month = "9",
    year = "2025"
}

@article{Tong:2025sui,
    author = "Tong, Hui and Elhatisari, Serdar and Mei{\ss}ner, Ulf-G.",
    title = "{Hyperneutron Stars from an Ab Initio Calculation}",
    eprint = "2502.14435",
    archivePrefix = "arXiv",
    primaryClass = "nucl-th",
    doi = "10.3847/1538-4357/adba47",
    journal = "Astrophys. J.",
    volume = "982",
    number = "2",
    pages = "164",
    year = "2025"
}

@article{Lonardoni:2014bwa,
    author = "Lonardoni, Diego and Lovato, Alessandro and Gandolfi, Stefano and Pederiva, Francesco",
    title = "{Hyperon Puzzle: Hints from Quantum Monte Carlo Calculations}",
    eprint = "1407.4448",
    archivePrefix = "arXiv",
    primaryClass = "nucl-th",
    reportNumber = "LA-UR-14-25265",
    doi = "10.1103/PhysRevLett.114.092301",
    journal = "Phys. Rev. Lett.",
    volume = "114",
    number = "9",
    pages = "092301",
    year = "2015"
}

@article{Ye:2024meg,
    author = "Ye, Jun-Ting and Wang, Rui and Wang, Si-Pei and Chen, Lie-Wen",
    title = "{High-density Symmetry Energy: A Key to the Solution of the Hyperon Puzzle}",
    eprint = "2411.18349",
    archivePrefix = "arXiv",
    primaryClass = "nucl-th",
    doi = "10.3847/1538-4357/add017",
    journal = "Astrophys. J.",
    volume = "985",
    number = "2",
    pages = "238",
    year = "2025"
}

@article{Muto:2024upf,
    author = "Muto, Takumi",
    title = "{Properties of a kaon-condensed phase in hyperon-mixed matter with three-baryon forces}",
    eprint = "2411.09967",
    archivePrefix = "arXiv",
    primaryClass = "nucl-th",
    doi = "10.1103/PhysRevC.111.045802",
    journal = "Phys. Rev. C",
    volume = "111",
    number = "4",
    pages = "045802",
    year = "2025"
}

@article{LiAng:2024xrl,
    author = "Ang, Li",
    title = "{Is There a {\textquotedblleft}hyperon puzzle{\textquotedblright} Problem in Neutron Star Study?}",
    doi = "10.11804/NuclPhysRev.41.QCS2023.09",
    journal = "Nucl. Phys. Rev.",
    volume = "41",
    number = "3",
    pages = "834--838",
    year = "2024"
}

@article{Tolos:2020aln,
    author = "Tolos, Laura and Fabbietti, Laura",
    title = "{Strangeness in Nuclei and Neutron Stars}",
    eprint = "2002.09223",
    archivePrefix = "arXiv",
    primaryClass = "nucl-ex",
    doi = "10.1016/j.ppnp.2020.103770",
    journal = "Prog. Part. Nucl. Phys.",
    volume = "112",
    pages = "103770",
    year = "2020"
}

@article{Tolos:2016hhl,
    author = "Tolos, Laura and Centelles, Mario and Ramos, Angels",
    title = "{Equation of State for Nucleonic and Hyperonic Neutron Stars with Mass and Radius Constraints}",
    eprint = "1610.00919",
    archivePrefix = "arXiv",
    primaryClass = "astro-ph.HE",
    doi = "10.3847/1538-4357/834/1/3",
    journal = "Astrophys. J.",
    volume = "834",
    number = "1",
    pages = "3",
    year = "2017"
}

@article{Chatterjee:2015pua,
    author = "Chatterjee, Debarati and Vida{\~n}a, Isaac",
    title = "{Do hyperons exist in the interior of neutron stars?}",
    eprint = "1510.06306",
    archivePrefix = "arXiv",
    primaryClass = "nucl-th",
    doi = "10.1140/epja/i2016-16029-x",
    journal = "Eur. Phys. J. A",
    volume = "52",
    number = "2",
    pages = "29",
    year = "2016"
}

@article{Gerstung:2020ktv,
    author = "Gerstung, Dominik and Kaiser, Norbert and Weise, Wolfram",
    title = "{Hyperon{\textendash}nucleon three-body forces and strangeness in neutron stars}",
    eprint = "2001.10563",
    archivePrefix = "arXiv",
    primaryClass = "nucl-th",
    doi = "10.1140/epja/s10050-020-00180-2",
    journal = "Eur. Phys. J. A",
    volume = "56",
    number = "6",
    pages = "175",
    year = "2020"
}

@article{Togashi:2016fky,
    author = "Togashi, H. and Hiyama, E. and Yamamoto, Y. and Takano, M.",
    title = "{Equation of state for neutron stars with hyperons by the variational method}",
    eprint = "1602.08106",
    archivePrefix = "arXiv",
    primaryClass = "nucl-th",
    doi = "10.1103/PhysRevC.93.035808",
    journal = "Phys. Rev. C",
    volume = "93",
    number = "3",
    pages = "035808",
    year = "2016"
}

@article{Miyatsu:2015kwa,
    author = "Miyatsu, Tsuyoshi and Cheoun, Myung-Ki and Saito, Koichi",
    title = "{Equation of State for Neutron Stars With Hyperons and Quarks in the Relativistic Hartree{\textendash}fock Approximation}",
    eprint = "1506.05552",
    archivePrefix = "arXiv",
    primaryClass = "nucl-th",
    doi = "10.1088/0004-637X/813/2/135",
    journal = "Astrophys. J.",
    volume = "813",
    number = "2",
    pages = "135",
    year = "2015"
}

@article{Sun:2022yor,
    author = "Sun, Xiangdong and Miao, Zhiqiang and Sun, Baoyuan and Li, Ang",
    title = "{Astrophysical Implications on Hyperon Couplings and Hyperon Star Properties with Relativistic Equations of States}",
    eprint = "2205.10631",
    archivePrefix = "arXiv",
    primaryClass = "astro-ph.HE",
    doi = "10.3847/1538-4357/ac9d9a",
    journal = "Astrophys. J.",
    volume = "942",
    number = "1",
    pages = "55",
    year = "2023"
}

@article{Sun:2018tmw,
    author = "Sun, Ting-Ting and Zhang, Shi-Sheng and Zhang, Qiu-Lan and Xia, Cheng-Jun",
    title = "{Strangeness and $\Delta$ resonance in compact stars with relativistic-mean-field models}",
    eprint = "1808.02207",
    archivePrefix = "arXiv",
    primaryClass = "nucl-th",
    doi = "10.1103/PhysRevD.99.023004",
    journal = "Phys. Rev. D",
    volume = "99",
    number = "2",
    pages = "023004",
    year = "2019"
}

@article{Yamamoto:2024mta,
    author = "Yamamoto, Y. and Yasutake, N. and Rijken, Th. A.",
    title = "{Nucleon-quark mixed matter and neutron-star equation~of state}",
    eprint = "2408.03812",
    archivePrefix = "arXiv",
    primaryClass = "nucl-th",
    doi = "10.1103/PhysRevC.110.025805",
    journal = "Phys. Rev. C",
    volume = "110",
    number = "2",
    pages = "025805",
    year = "2024"
}

@article{Abbott:2023coj,
    author = "Abbott, Ryan and Detmold, William and Romero-L{\'o}pez, Fernando and Davoudi, Zohreh and Illa, Marc and Parre{\~n}o, Assumpta and Perry, Robert J. and Shanahan, Phiala E. and Wagman, Michael L.",
    collaboration = "NPLQCD",
    title = "{Lattice quantum chromodynamics at large isospin density}",
    eprint = "2307.15014",
    archivePrefix = "arXiv",
    primaryClass = "hep-lat",
    reportNumber = "MIT-CTP/5560, UMD-PP-023-03, FERMILAB-PUB-23-382-T",
    doi = "10.1103/PhysRevD.108.114506",
    journal = "Phys. Rev. D",
    volume = "108",
    number = "11",
    pages = "114506",
    year = "2023"
}

@article{Abbott:2024vhj,
    author = "Abbott, Ryan and Detmold, William and Illa, Marc and Parre{\~n}o, Assumpta and Perry, Robert J. and Romero-L{\'o}pez, Fernando and Shanahan, Phiala E. and Wagman, Michael L.",
    collaboration = "NPLQCD",
    title = "{QCD Constraints on Isospin-Dense Matter and the Nuclear Equation of State}",
    eprint = "2406.09273",
    archivePrefix = "arXiv",
    primaryClass = "hep-lat",
    reportNumber = "MIT-CTP/5730,FERMILAB-PUB-24-0333-T, MIT-CTP/5729, FERMILAB-PUB-24-0333-T",
    doi = "10.1103/PhysRevLett.134.011903",
    journal = "Phys. Rev. Lett.",
    volume = "134",
    number = "1",
    pages = "011903",
    year = "2025"
}

@article{Brandt:2022hwy,
    author = "Brandt, Bastian B. and Cuteri, Francesca and Endrodi, Gergely",
    title = "{Equation of state and speed of sound of isospin-asymmetric QCD on the lattice}",
    eprint = "2212.14016",
    archivePrefix = "arXiv",
    primaryClass = "hep-lat",
    doi = "10.1007/JHEP07(2023)055",
    journal = "JHEP",
    volume = "07",
    pages = "055",
    year = "2023"
}

@article{Brandt:2017oyy,
    author = "Brandt, B. B. and Endrodi, G. and Schmalzbauer, S.",
    title = "{QCD phase diagram for nonzero isospin-asymmetry}",
    eprint = "1712.08190",
    archivePrefix = "arXiv",
    primaryClass = "hep-lat",
    doi = "10.1103/PhysRevD.97.054514",
    journal = "Phys. Rev. D",
    volume = "97",
    number = "5",
    pages = "054514",
    year = "2018"
}

@article{Iida:2020emi,
    author = "Iida, Kei and Itou, Etsuko and Lee, Tong-Gyu",
    title = "{Relative scale setting for two-color QCD with $N_f$=2 Wilson fermions}",
    eprint = "2008.06322",
    archivePrefix = "arXiv",
    primaryClass = "hep-lat",
    doi = "10.1093/ptep/ptaa170",
    journal = "PTEP",
    volume = "2021",
    number = "1",
    pages = "013B05",
    year = "2021"
}

@misc{Itou:2025vcy,
    author = "Itou, Etsuko",
    title = "{Lattice results for the equation of state in dense QCD-like theories}",
    eprint = "2508.03090",
    archivePrefix = "arXiv",
    primaryClass = "hep-lat",
    reportNumber = "YITP-25-117, RIKEN-iTHEMS-Report-25",
    month = "8",
    year = "2025"
}

@article{Strodthoff:2013cua,
    author = "Strodthoff, Nils and von Smekal, Lorenz",
    title = "{Polyakov-Quark-Meson-Diquark Model for two-color QCD}",
    eprint = "1306.2897",
    archivePrefix = "arXiv",
    primaryClass = "hep-ph",
    doi = "10.1016/j.physletb.2014.03.008",
    journal = "Phys. Lett. B",
    volume = "731",
    pages = "350--357",
    year = "2014"
}

@article{Ivanytskyi:2025cnn,
    author = "Ivanytskyi, Oleksii",
    title = "{Quarkyonic picture of isospin QCD}",
    eprint = "2505.07076",
    archivePrefix = "arXiv",
    primaryClass = "nucl-th",
    doi = "10.1103/831v-8mp4",
    journal = "Phys. Rev. D",
    volume = "112",
    number = "3",
    pages = "034001",
    year = "2025"
}

@article{Kojo:2021ugu,
    author = "Kojo, Toru",
    title = "{Stiffening of matter in quark-hadron continuity}",
    eprint = "2106.06687",
    archivePrefix = "arXiv",
    primaryClass = "nucl-th",
    doi = "10.1103/PhysRevD.104.074005",
    journal = "Phys. Rev. D",
    volume = "104",
    number = "7",
    pages = "074005",
    year = "2021"
}

@article{Kojo:2021hqh,
    author = "Kojo, Toru and Suenaga, Daiki",
    title = "{Peaks of sound velocity in two color dense QCD: Quark saturation effects and semishort range correlations}",
    eprint = "2110.02100",
    archivePrefix = "arXiv",
    primaryClass = "hep-ph",
    doi = "10.1103/PhysRevD.105.076001",
    journal = "Phys. Rev. D",
    volume = "105",
    number = "7",
    pages = "076001",
    year = "2022"
}

@article{Fujimoto:2023mzy,
    author = "Fujimoto, Yuki and Kojo, Toru and McLerran, Larry D.",
    title = "{Momentum Shell in Quarkyonic Matter from Explicit Duality: A Dual Model for Cold, Dense QCD}",
    eprint = "2306.04304",
    archivePrefix = "arXiv",
    primaryClass = "nucl-th",
    reportNumber = "INT-PUB-23-018",
    doi = "10.1103/PhysRevLett.132.112701",
    journal = "Phys. Rev. Lett.",
    volume = "132",
    number = "11",
    pages = "112701",
    year = "2024"
}

@article{Gao:2024jlp,
    author = "Gao, Bikai and Harada, Masayasu",
    title = "{Quarkyonic matter with chiral symmetry restoration}",
    eprint = "2410.16649",
    archivePrefix = "arXiv",
    primaryClass = "nucl-th",
    doi = "10.1103/PhysRevD.111.016024",
    journal = "Phys. Rev. D",
    volume = "111",
    number = "1",
    pages = "016024",
    year = "2025"
}

@misc{Gao:2025okn,
    author = "Gao, Bikai and Marczenko, Micha{\l}",
    title = "{Suppression of dynamical momentum-space shell by chiral symmetry}",
    eprint = "2509.03138",
    archivePrefix = "arXiv",
    primaryClass = "nucl-th",
    month = "9",
    year = "2025"
}

@article{Bentz:2025rla,
    author = {Bentz, Wolfgang and Clo{\"e}t, Ian C.},
    title = "{Effects of Quark Core Sizes of Baryons in Neutron Star Matter}",
    eprint = "2503.20564",
    archivePrefix = "arXiv",
    primaryClass = "nucl-th",
    doi = "10.3390/sym17040505",
    journal = "Symmetry",
    volume = "17",
    number = "4",
    pages = "505",
    year = "2025"
}

@misc{Gartlein:2025zhd,
    author = {G{\"a}rtlein, Christoph and Ivanytskyi, Oleksii and Sagun, Violetta and Lopes, Il{\'\i}dio},
    title = "{Color-superconducting quarkyonic matter}",
    eprint = "2509.03517",
    archivePrefix = "arXiv",
    primaryClass = "nucl-th",
    month = "9",
    year = "2025"
}

@misc{Nikolakopoulos:2025fgw,
    author = "Nikolakopoulos, Alexis and Miller, Gerald A.",
    title = "{Quark Phase Space Distributions in Nuclei}",
    eprint = "2506.22670",
    archivePrefix = "arXiv",
    primaryClass = "nucl-th",
    reportNumber = "NT@UW-25-7",
    month = "6",
    year = "2025"
}

@article{McLerran:2024rvk,
    author = "McLerran, Larry and Miller, Gerald A.",
    title = "{Quark Pauli principle and the transmutation of nuclear matter}",
    eprint = "2405.11074",
    archivePrefix = "arXiv",
    primaryClass = "nucl-th",
    reportNumber = "NT@UW-24-07",
    doi = "10.1103/PhysRevC.110.045203",
    journal = "Phys. Rev. C",
    volume = "110",
    number = "4",
    pages = "045203",
    year = "2024"
}

@article{Koch:2024qnz,
    author = "Koch, Volker and McLerran, Larry and Miller, Gerald A. and Vovchenko, Volodymyr",
    title = "{Examining the possibility that normal nuclear matter is quarkyonic}",
    eprint = "2403.15375",
    archivePrefix = "arXiv",
    primaryClass = "nucl-th",
    doi = "10.1103/PhysRevC.110.025201",
    journal = "Phys. Rev. C",
    volume = "110",
    number = "2",
    pages = "025201",
    year = "2024"
}

@article{Togashi:2017mjp,
    author = "Togashi, H. and Nakazato, K. and Takehara, Y. and Yamamuro, S. and Suzuki, H. and Takano, M.",
    title = "{Nuclear equation of state for core-collapse supernova simulations with realistic nuclear forces}",
    eprint = "1702.05324",
    archivePrefix = "arXiv",
    primaryClass = "nucl-th",
    doi = "10.1016/j.nuclphysa.2017.02.010",
    journal = "Nucl. Phys. A",
    volume = "961",
    pages = "78--105",
    year = "2017"
}

@article{Kojo:2024sca,
    author = "Kojo, Toru and Suenaga, Daiki and Chiba, Ryuji",
    title = "{Isospin QCD as a Laboratory for Dense QCD}",
    eprint = "2406.11059",
    archivePrefix = "arXiv",
    primaryClass = "hep-ph",
    doi = "10.3390/universe10070293",
    journal = "Universe",
    volume = "10",
    number = "7",
    pages = "293",
    year = "2024"
}

@article{Chiba:2023ftg,
    author = "Chiba, Ryuji and Kojo, Toru",
    title = "{Sound velocity peak and conformality in isospin QCD}",
    eprint = "2304.13920",
    archivePrefix = "arXiv",
    primaryClass = "hep-ph",
    doi = "10.1103/PhysRevD.109.076006",
    journal = "Phys. Rev. D",
    volume = "109",
    number = "7",
    pages = "076006",
    year = "2024"
}

@article{Drischler:2021bup,
	archiveprefix = {arXiv},
	author = {Drischler, Christian and Han, Sophia and Reddy, Sanjay},
	doi = {10.1103/PhysRevC.105.035808},
	eprint = {2110.14896},
	journal = {Phys. Rev. C},
	number = {3},
	pages = {035808},
	primaryclass = {nucl-th},
	reportnumber = {INT-PUB-21-018, N3AS-21-011},
	title = {{Large and massive neutron stars: Implications for the sound speed within QCD of dense matter}},
	volume = {105},
	year = {2022}}

@article{Ma:2019ery,
	archiveprefix = {arXiv},
	author = {Ma, Yong-Liang and Rho, Mannque},
	doi = {10.1016/j.ppnp.2020.103791},
	eprint = {1909.05889},
	journal = {Prog. Part. Nucl. Phys.},
	pages = {103791},
	primaryclass = {nucl-th},
	title = {{Towards the hadron\textendash{}quark continuity via a topology change in compact stars}},
	volume = {113},
	year = {2020}}

@misc{Marczenko:2022jhl,
	archiveprefix = {arXiv},
	author = {Marczenko, Micha\l{} and McLerran, Larry and Redlich, Krzysztof and Sasaki, Chihiro},
	eprint = {2207.13059},
	month = {7},
	primaryclass = {nucl-th},
	title = {{Reaching percolation and conformal limits in neutron stars}},
	year = {2022}}

@article{Adhikari:2018cea,
	archiveprefix = {arXiv},
	author = {Adhikari, Prabal and Andersen, Jens O. and Kneschke, Patrick},
	doi = {10.1103/PhysRevD.98.074016},
	eprint = {1805.08599},
	journal = {Phys. Rev. D},
	number = {7},
	pages = {074016},
	primaryclass = {hep-ph},
	title = {{Pion condensation and phase diagram in the Polyakov-loop quark-meson model}},
	volume = {98},
	year = {2018}}

@article{Adhikari:2016eef,
	archiveprefix = {arXiv},
	author = {Adhikari, Prabal and Andersen, Jens O. and Kneschke, Patrick},
	doi = {10.1103/PhysRevD.95.036017},
	eprint = {1612.03668},
	journal = {Phys. Rev. D},
	number = {3},
	pages = {036017},
	primaryclass = {hep-ph},
	title = {{On-shell parameter fixing in the quark-meson model}},
	volume = {95},
	year = {2017}}

@misc{Ayala:2023cnt,
	archiveprefix = {arXiv},
	author = {Ayala, Alejandro and Bandyopadhyay, Aritra and Farias, Ricardo L. S. and Hern\'andez, Luis A. and Hern\'andez, Jos\'e Luis},
	eprint = {2301.13633},
	month = {1},
	primaryclass = {hep-ph},
	title = {{QCD equation of state at finite isospin density from the linear sigma model with quarks: The cold case}},
	year = {2023}}

@article{Strodthoff:2011tz,
	archiveprefix = {arXiv},
	author = {Strodthoff, Nils and Schaefer, Bernd-Jochen and von Smekal, Lorenz},
	doi = {10.1103/PhysRevD.85.074007},
	eprint = {1112.5401},
	journal = {Phys. Rev. D},
	pages = {074007},
	primaryclass = {hep-ph},
	title = {{Quark-meson-diquark model for two-color QCD}},
	volume = {85},
	year = {2012}}

@article{Iida:2022hyy,
    author = "Iida, Kei and Itou, Etsuko",
    title = "{Velocity of sound beyond the high-density relativistic limit from lattice simulation of dense two-color QCD}",
    eprint = "2207.01253",
    archivePrefix = "arXiv",
    primaryClass = "hep-ph",
    reportNumber = "RIKEN-iTHEMS-Report-22",
    doi = "10.1093/ptep/ptac137",
    journal = "PTEP",
    volume = "2022",
    number = "11",
    pages = "111B01",
    year = "2022"
}

@article{Murakami:2022lmq,
    author = "Murakami, Kotaro and Suenaga, Daiki and Iida, Kei and Itou, Etsuko",
    title = "{Measurement of hadron masses in 2-color finite density QCD}",
    eprint = "2211.13472",
    archivePrefix = "arXiv",
    primaryClass = "hep-lat",
    reportNumber = "YITP-22-146, RIKEN-iTHEMS-Report-22",
    doi = "10.22323/1.430.0154",
    journal = "PoS",
    volume = "LATTICE2022",
    pages = "154",
    year = "2023"
}

@article{Iida:2019rah,
	archiveprefix = {arXiv},
	author = {Iida, Kei and Itou, Etsuko and Lee, Tong-Gyu},
	doi = {10.1007/JHEP01(2020)181},
	eprint = {1910.07872},
	journal = {JHEP},
	pages = {181},
	primaryclass = {hep-lat},
	title = {{Two-colour QCD phases and the topology at low temperature and high density}},
	volume = {01},
	year = {2020}}

@article{Astrakhantsev:2020tdl,
	archiveprefix = {arXiv},
	author = {Astrakhantsev, N. and Braguta, V.V. and Ilgenfritz, E.M. and Kotov, A.Yu. and Nikolaev, A.A.},
	doi = {10.1103/PhysRevD.102.074507},
	eprint = {2007.07640},
	journal = {Phys. Rev. D},
	number = {7},
	pages = {074507},
	primaryclass = {hep-lat},
	title = {{Lattice study of thermodynamic properties of dense QC$_2$D}},
	volume = {102},
	year = {2020}}

@article{Astrakhantsev:2018uzd,
	archiveprefix = {arXiv},
	author = {Astrakhantsev, N. Yu. and Bornyakov, V. G. and Braguta, V. V. and Ilgenfritz, E. -M. and Kotov, A. Yu. and Nikolaev, A. A. and Rothkopf, A.},
	doi = {10.1007/JHEP05(2019)171},
	eprint = {1808.06466},
	journal = {JHEP},
	pages = {171},
	primaryclass = {hep-lat},
	title = {{Lattice study of static quark-antiquark interactions in dense quark matter}},
	volume = {05},
	year = {2019}}

@article{Bornyakov:2020kyz,
	archiveprefix = {arXiv},
	author = {Bornyakov, V. G. and Braguta, V. V. and Nikolaev, A. A. and Rogalyov, R. N.},
	doi = {10.1103/PhysRevD.102.114511},
	eprint = {2003.00232},
	journal = {Phys. Rev. D},
	pages = {114511},
	primaryclass = {hep-lat},
	title = {{Effects of Dense Quark Matter on Gluon Propagators in Lattice QC$_2$D}},
	volume = {102},
	year = {2020}}

@article{Boz:2018crd,
	archiveprefix = {arXiv},
	author = {Boz, Tamer and Hajizadeh, Ouraman and Maas, Axel and Skullerud, Jon-Ivar},
	doi = {10.1103/PhysRevD.99.074514},
	eprint = {1812.08517},
	journal = {Phys. Rev. D},
	number = {7},
	pages = {074514},
	primaryclass = {hep-lat},
	title = {{Finite-density gauge correlation functions in QC2D}},
	volume = {99},
	year = {2019}}

@article{Bornyakov:2017txe,
	archiveprefix = {arXiv},
	author = {Bornyakov, V. G. and Braguta, V. V. and Ilgenfritz, E. -M. and Kotov, A. Yu. and Molochkov, A. V. and Nikolaev, A. A.},
	doi = {10.1007/JHEP03(2018)161},
	eprint = {1711.01869},
	journal = {JHEP},
	pages = {161},
	primaryclass = {hep-lat},
	title = {{Observation of deconfinement in a cold dense quark medium}},
	volume = {03},
	year = {2018}}

@article{Braguta:2016cpw,
	archiveprefix = {arXiv},
	author = {Braguta, V. V. and Ilgenfritz, E. -M. and Kotov, A. Yu. and Molochkov, A. V. and Nikolaev, A. A.},
	doi = {10.1103/PhysRevD.94.114510},
	eprint = {1605.04090},
	journal = {Phys. Rev. D},
	number = {11},
	pages = {114510},
	primaryclass = {hep-lat},
	title = {{Study of the phase diagram of dense two-color QCD within lattice simulation}},
	volume = {94},
	year = {2016}}

@article{Braguta:2015zta,
	archiveprefix = {arXiv},
	author = {Braguta, V. V. and Goy, V. A. and Ilgenfritz, E. -M. and Kotov, A. Yu. and Molochkov, A. V. and Muller-Preussker, M. and Petersson, B.},
	doi = {10.1007/JHEP06(2015)094},
	eprint = {1503.06670},
	journal = {JHEP},
	pages = {094},
	primaryclass = {hep-lat},
	title = {{Two-Color QCD with Non-zero Chiral Chemical Potential}},
	volume = {06},
	year = {2015}}

@article{Braguta:2014gea,
	archiveprefix = {arXiv},
	author = {Braguta, V. and Chernodub, M. N. and Goy, V. A. and Landsteiner, K. and Molochkov, A. V. and Polikarpov, M. I.},
	doi = {10.1103/PhysRevD.89.074510},
	eprint = {1401.8095},
	journal = {Phys. Rev. D},
	number = {7},
	pages = {074510},
	primaryclass = {hep-lat},
	reportnumber = {IFT-UAM-CSIC-14-002},
	title = {{Temperature dependence of the axial magnetic effect in two-color quenched QCD}},
	volume = {89},
	year = {2014}}

@misc{Buividovich:2020gnl,
	archiveprefix = {arXiv},
	author = {Buividovich, P. V. and Smith, D. and von Smekal, L.},
	eprint = {2012.05184},
	month = {12},
	primaryclass = {hep-lat},
	title = {{Numerical Study of the Chiral Separation Effect in Two-Color QCD at Finite Density}},
	year = {2020}}

@book{schrieffer1999theory,
	author = {Schrieffer, J.R.},
	isbn = {9780738201207},
	lccn = {99060035},
	publisher = {Avalon Publishing},
	series = {Advanced Books Classics},
	title = {Theory Of Superconductivity},
	url = {https://books.google.co.jp/books?id=let7wRir74MC},
	year = {1999}}

@inbook{Leggett_book,
	author = {Leggett, A. J. and S. Zhang},
	booktitle = {The BCS-BEC Crossover and the Unitary Fermi Gas},
	doi = {10.1007/978-3-642-21978-8_2},
	editor = {Wilhelm Zwerger},
	isbn = {9783642219771},
	pages = {33--47},
	series = {Lecture Notes in Physics},
	title = {The BEC-BCS crossover: Some history and some general observations},
	year = {2012}}

@inbook{BCS-BEC_Parish,
	adsurl = {https://ui.adsabs.harvard.edu/abs/2015qgee.book..179P},
	author = {{Parish}, Meera M.},
	doi = {10.1142/9781783264766\_0009},
	pages = {179-197},
	title = {{The BCS-BEC Crossover, Quantum Gas Experiments: Exploring Many-Body States. Edited by TORMA PAIVI ET AL. (World Scientific Publishing Co. Pte. Ltd, 2015)}},
	year = {2015}}

@article{Demorest:2010bx,
	archiveprefix = {arXiv},
	author = {Demorest, Paul and Pennucci, Tim and Ransom, Scott and Roberts, Mallory and Hessels, Jason},
	doi = {10.1038/nature09466},
	eprint = {1010.5788},
	journal = {Nature},
	pages = {1081--1083},
	primaryclass = {astro-ph.HE},
	title = {{Shapiro Delay Measurement of A Two Solar Mass Neutron Star}},
	volume = {467},
	year = {2010}}

@article{Fonseca:2021wxt,
	archiveprefix = {arXiv},
	author = {Fonseca, E. and others},
	doi = {10.3847/2041-8213/ac03b8},
	eprint = {2104.00880},
	journal = {Astrophys. J. Lett.},
	number = {1},
	pages = {L12},
	primaryclass = {astro-ph.HE},
	title = {{Refined Mass and Geometric Measurements of the High-mass PSR J0740+6620}},
	volume = {915},
	year = {2021}}

@article{Antoniadis:2013pzd,
	archiveprefix = {arXiv},
	author = {Antoniadis, John and others},
	doi = {10.1126/science.1233232},
	eprint = {1304.6875},
	journal = {Science},
	pages = {6131},
	primaryclass = {astro-ph.HE},
	title = {{A Massive Pulsar in a Compact Relativistic Binary}},
	volume = {340},
	year = {2013}}

@article{TheLIGOScientific:2017qsa,
	archiveprefix = {arXiv},
	author = {Abbott, B.P. and others},
	collaboration = {LIGO Scientific, Virgo},
	doi = {10.1103/PhysRevLett.119.161101},
	eprint = {1710.05832},
	journal = {Phys. Rev. Lett.},
	number = {16},
	pages = {161101},
	primaryclass = {gr-qc},
	reportnumber = {LIGO-P170817},
	title = {{GW170817: Observation of Gravitational Waves from a Binary Neutron Star Inspiral}},
	volume = {119},
	year = {2017}}

@article{Kojo:2014rca,
	archiveprefix = {arXiv},
	author = {Kojo, Toru and Powell, Philip D. and Song, Yifan and Baym, Gordon},
	doi = {10.1103/PhysRevD.91.045003},
	eprint = {1412.1108},
	journal = {Phys. Rev. D},
	number = {4},
	pages = {045003},
	primaryclass = {hep-ph},
	title = {{Phenomenological QCD equation of state for massive neutron stars}},
	volume = {91},
	year = {2015}}

@article{Baym:2017whm,
	archiveprefix = {arXiv},
	author = {Baym, Gordon and Hatsuda, Tetsuo and Kojo, Toru and Powell, Philip D. and Song, Yifan and Takatsuka, Tatsuyuki},
	doi = {10.1088/1361-6633/aaae14},
	eprint = {1707.04966},
	journal = {Rept. Prog. Phys.},
	number = {5},
	pages = {056902},
	primaryclass = {astro-ph.HE},
	reportnumber = {RIKEN-ITHEMS-REPORT-17, RIKEN-QHP-316, RIKEN-iTHEMS-Report-17},
	title = {{From hadrons to quarks in neutron stars: a review}},
	volume = {81},
	year = {2018}}

@article{Masuda:2012kf,
	archiveprefix = {arXiv},
	author = {Masuda, Kota and Hatsuda, Tetsuo and Takatsuka, Tatsuyuki},
	doi = {10.1088/0004-637X/764/1/12},
	eprint = {1205.3621},
	journal = {Astrophys. J.},
	pages = {12},
	primaryclass = {nucl-th},
	title = {{Hadron-Quark Crossover and Massive Hybrid Stars with Strangeness}},
	volume = {764},
	year = {2013}}

@article{Masuda:2012ed,
	archiveprefix = {arXiv},
	author = {Masuda, Kota and Hatsuda, Tetsuo and Takatsuka, Tatsuyuki},
	doi = {10.1093/ptep/ptt045},
	eprint = {1212.6803},
	journal = {PTEP},
	number = {7},
	pages = {073D01},
	primaryclass = {nucl-th},
	title = {{Hadron--quark crossover and massive hybrid stars}},
	volume = {2013},
	year = {2013}}

@misc{Saito:2025yld,
    author = "Saito, Koichi and Miyatsu, Tsuyoshi and Cheoun, Myung-Ki",
    title = "{A Quarkyonic Quark-Meson Coupling Model for Nuclear and Neutron Matter}",
    eprint = "2512.04505",
    archivePrefix = "arXiv",
    primaryClass = "nucl-th",
    month = "12",
    year = "2025"
}

@article{McLerran:2018hbz,
	archiveprefix = {arXiv},
	author = {McLerran, Larry and Reddy, Sanjay},
	doi = {10.1103/PhysRevLett.122.122701},
	eprint = {1811.12503},
	journal = {Phys. Rev. Lett.},
	number = {12},
	pages = {122701},
	primaryclass = {nucl-th},
	reportnumber = {INT-PUB-18-060},
	title = {{Quarkyonic Matter and Neutron Stars}},
	volume = {122},
	year = {2019}}

@article{Jeong:2019lhv,
	archiveprefix = {arXiv},
	author = {Jeong, Kie Sang and McLerran, Larry and Sen, Srimoyee},
	doi = {10.1103/PhysRevC.101.035201},
	eprint = {1908.04799},
	journal = {Phys. Rev. C},
	number = {3},
	pages = {035201},
	primaryclass = {nucl-th},
	reportnumber = {INT-PUB-19-048, APCTP Pre 2019-020},
	title = {{Dynamically generated momentum space shell structure of quarkyonic matter via an excluded volume model}},
	volume = {101},
	year = {2020}}

@article{Duarte:2020xsp,
	archiveprefix = {arXiv},
	author = {Duarte, Dyana C. and Hernandez-Ortiz, Saul and Jeong, Kie Sang},
	doi = {10.1103/PhysRevC.102.025203},
	eprint = {2003.02362},
	journal = {Phys. Rev. C},
	number = {2},
	pages = {025203},
	primaryclass = {nucl-th},
	reportnumber = {INT-PUB-20-001},
	title = {{Excluded-volume model for quarkyonic Matter: Three-flavor baryon-quark Mixture}},
	volume = {102},
	year = {2020}}

@article{Duarte:2020kvi,
	archiveprefix = {arXiv},
	author = {Duarte, Dyana C. and Hernandez-Ortiz, Saul and Jeong, Kie Sang},
	doi = {10.1103/PhysRevC.102.065202},
	eprint = {2007.08098},
	journal = {Phys. Rev. C},
	number = {6},
	pages = {065202},
	primaryclass = {nucl-th},
	reportnumber = {INT-PUB-20-028},
	title = {{Excluded-volume model for quarkyonic matter. II. Three-flavor shell-like distribution of baryons in phase space}},
	volume = {102},
	year = {2020}}

@article{Zhao:2020dvu,
	archiveprefix = {arXiv},
	author = {Zhao, Tianqi and Lattimer, James M.},
	doi = {10.1103/PhysRevD.102.023021},
	eprint = {2004.08293},
	journal = {Phys. Rev. D},
	number = {2},
	pages = {023021},
	primaryclass = {astro-ph.HE},
	title = {{Quarkyonic Matter Equation of State in Beta-Equilibrium}},
	volume = {102},
	year = {2020}}

@article{McLerran:2007qj,
	archiveprefix = {arXiv},
	author = {McLerran, Larry and Pisarski, Robert D.},
	doi = {10.1016/j.nuclphysa.2007.08.013},
	eprint = {0706.2191},
	journal = {Nucl. Phys. A},
	pages = {83--100},
	primaryclass = {hep-ph},
	title = {{Phases of cold, dense quarks at large N(c)}},
	volume = {796},
	year = {2007}}

@article{Tajima:2024qzj,
    author = "Tajima, Hiroyuki and Iida, Kei and Kojo, Toru and Liang, Haozhao",
    title = "{Tripling Fluctuations and Peaked Sound Speed in Fermionic Matter}",
    eprint = "2412.04971",
    archivePrefix = "arXiv",
    primaryClass = "hep-ph",
    reportNumber = "RIKEN-iTHEMS-Report-24",
    doi = "10.1103/4ywp-752m",
    journal = "Phys. Rev. Lett.",
    volume = "135",
    number = "4",
    pages = "042701",
    year = "2025"
}

@misc{Fujimoto:2024doc,
    author = "Fujimoto, Yuki and Kojo, Toru and McLerran, Larry",
    title = "{Quarkyonic matter pieces together the hyperon puzzle}",
    eprint = "2410.22758",
    archivePrefix = "arXiv",
    primaryClass = "nucl-th",
    reportNumber = "INT-PUB-24-056, RIKEN-iTHEMS-Report-24",
    month = "10",
    year = "2024"
}

@article{Fukushima:2020cmk,
	archiveprefix = {arXiv},
	author = {Fukushima, Kenji and Kojo, Toru and Weise, Wolfram},
	doi = {10.1103/PhysRevD.102.096017},
	eprint = {2008.08436},
	journal = {Phys. Rev. D},
	number = {9},
	pages = {096017},
	primaryclass = {hep-ph},
	title = {{Hard-core deconfinement and soft-surface delocalization from nuclear to quark matter}},
	volume = {102},
	year = {2020}}

@article{Takahashi:2009ef,
	archiveprefix = {arXiv},
	author = {Takahashi, Toru T. and Kanada-En'yo, Yoshiko},
	doi = {10.1103/PhysRevD.82.094506},
	eprint = {0912.0691},
	journal = {Phys. Rev. D},
	pages = {094506},
	primaryclass = {hep-lat},
	title = {{Hadron-hadron interaction from SU(2) lattice QCD}},
	volume = {82},
	year = {2010}}

\end{document}